\documentclass[showpacs,reprint,nofootinbib,longbibliography,superscriptaddress]{revtex4-1}

\usepackage{hyperref}
\usepackage[utf8x]{inputenc}
\usepackage{epsfig,dsfont}
\usepackage{color}
\usepackage{tikz}
\usepackage[normalem]{ulem}
\usepackage{color}
\usepackage{verbatim}

\usepackage{amsmath}
\usepackage{amsfonts}
\usepackage{amssymb}

\newcommand{\ph}{\varphi}
\newcommand{\sti}{\sigma}

\newcommand{\dsc}{r}

\newcommand{\be}{\begin{equation}}
\newcommand{\ee}{\end{equation}}

\begin{document}

\title{The two-dimensional O(3) model at nonzero density:\\
from dual lattice simulations to repulsive bosons}

\author{Falk Bruckmann}
\affiliation{Universit\"at Regensburg, Institut f\"ur Physik, Universit\"atstra{\ss}e 31, 93053 Regensburg, Germany}
  
\author{Christof Gattringer}
\affiliation{Universit\"at Graz, Institut f\"ur Physik, Universit\"atsplatz 5, 8010 Graz, Austria}

\author{Thomas Kloiber}
\affiliation{Universit\"at Regensburg, Institut f\"ur Physik, Universit\"atstra{\ss}e 31, 93053 Regensburg, Germany}
\affiliation{Universit\"at Graz, Institut f\"ur Physik, Universit\"atsplatz 5, 8010 Graz, Austria}

\author{and Tin Sulejmanpasic}
\affiliation{North Carolina State University, Department of Physics, Raleigh, NC 27695-8202, USA}

\begin{abstract}
We discuss the thermodynamics of the O(3) nonlinear sigma model in 1+1 dimensions at nonzero chemical potential (equivalent to a magnetic field). In its conventional field theory representation the model suffers from a sign problem. By dualizing the lattice model we present, for the first time, nonzero density data of an asymptotically free theory with dynamical mass-gap. We find a quantum phase transition at zero temperature where as a function of the chemical potential the density  assumes a nonzero value. Measuring the spin stiffness we present evidence for a corresponding dynamical critical exponent $z$ close to $2$. The low energy O(3) model is conjectured to be described by a massive boson triplet with repulsive interactions. We confirm the universal square root behavior expected for such a system at low density (and temperature) and compare our data to the results of Bethe ansatz solutions of the relativistic and non-relativistic one-dimensional Bose gas. We also comment on a potential Berezinskii-
Kosterlitz-Thouless transition at nonzero density.
\end{abstract}

%\date
\pacs{11.15.Ha,12.40.Ee}

\maketitle

\section{Introduction}

Analyzing quantum field theories at nonzero temperature and chemical potential is of interest not only for studying their 
thermodynamical properties, but may also provide deep insights into the physical structure of the theory under 
consideration. A recent such example is given in \cite{BruGaKloSu_2}, where we demonstrated 
that the grandcanonical ensemble at low temperature and small volumes can be connected to scattering data.

Before we outline the physics results presented in this paper, we briefly mention the challenges of thermodynamics for 
numerical lattice simulations, one of our best ab-initio tools (since analytic approaches are typically limited). Monte Carlo 
simulations are based on importance sampling and, while finite temperature simulations are routine, simulations at 
nonzero chemical potentials $\mu$ are in many cases plagued by the complex action/sign problem: 
at nonzero $\mu$ the 
action $S$ has a non-vanishing imaginary part and the Boltzmann factor $e^{-S}$ cannot be used as a probability weight 
in a stochastic process. One of the most successful methods is to employ dualization of the lattice path integral to new, 
so-called `dual variables', where the partition sum has only real and positive contributions (see, e.g., 
\cite{review1,review2} for reviews). Although, it is not completely clear for which classes of 
models a dual representation is useful 
for treating the sign problem, in models where it is successful the dual representation has allowed the exploration of the 
finite density phase diagrams, and the corresponding data also serve as a benchmark for 
other approaches to the complex action problem.

We would like to stress that dual representations not only enable simulations at nonzero density, 
but also reveal physical aspects  complementary to those of the standard representation. 
In the dual representation the dynamical degrees of freedom are 
worldlines and the chemical potential couples to their temporal component. Via a discrete version of 
current conservation on a space-time lattice one finds that the chemical potential indeed couples to the 
temporal winding number of the worldlines. Thus the net number of particles 
(charged under the Noether charge of the corresponding symmetry) can be identified with the temporal net winding 
number of the worldlines and the Noether charge becomes topologically conserved.
As such the worldlines  winding around the time direction have a direct interpretation as quantum states carrying the 
corresponding quantum number. Since this is an all-scale statement, the dual worldlines carry direct information about 
the infrared physics, generally obscured in the conventional formulation of asymptotically free theories. 

The O(3) nonlinear sigma model, which we are dealing with in this work, is conjectured to possess a particle triplet as its 
infrared excitations, the mass of which is generated dynamically. The dual wave function method of \cite{BruGaKloSu_2} 
indeed utilizes the spatial distance of the worldlines to obtain information about the particles' scattering. In the case of 
Quantum Chromodynamics (QCD), which is also asymptotically free, the low energy excitations are not the colored 
quarks and gluons, but colorless hadrons, with a large fraction of their masses being dynamically generated. A related 
dual lattice representation is that in terms of meson hoppings plus (anti-)baryon worldlines, to which the baryon 
chemical potential couples \cite{Rossi:1984cv,Karsch:1988zx} (this dual representation, however, does not 
remove the sign problem completely and does not take into account the gauge action).

To be more concrete, we introduce a chemical potential for an O(2) subgroup of the O(3) symmetry of the sigma model 
in two dimensions.  At low temperatures a nonzero density is expected to be induced into the system when $\mu$ 
reaches the threshold of the particle mass. In the condensed matter context $\mu$ can be viewed as a constant 
magnetic field inducing a magnetization; see Secs.~11 and 19 of \cite{Sachdev}. We will analyze this transition in detail 
extending our previous simulations to large volumes. Beyond the transition the system explicitly breaks the global 
internal symmetry from $O(3)$ to $O(2)$ and in principle allows classical vortex solutions with half-integer topological 
charge \cite{falktin}. Although at asymptotically large densities vortices are expected to bind in neutral pairs, it was 
conjectured in \cite{falktin} that these vortices will be liberated at small densities, so that a 
Berezinskii-Kosterlitz-Thouless (BKT) transition \cite{Berezinsky,KosterlitzThouless} 
may happen at some finite value of the density.

Our main findings are twofold: Concerning the O(3) phase diagram we find a threshold \textit{crossover} at nonzero 
temperatures and a  \textit{quantum phase transition} at zero temperature, as a function of $\mu$ at the mass threshold. 
Note that lattice simulations become expensive in this limit, as both temporal and spatial extension must be taken to 
infinity. Using the concept of spin stiffness we analyze spatial correlations and present evidence for a dynamical critical 
exponent $z$ close to $2$, which is consistent with the non-relativistic free fermions to which a model at low density belongs to. 
We do not directly see 
signatures of a BKT transition conjectured in \cite{falktin} in any of the observables studied in this work.

Secondly, the numerical data for the density near the transition can be described by a square root. The latter is universal 
for \textit{one-dimensional repulsive bosons}. The simplest example of which is the 
non-relativistic quantum particle gas with repulsive delta-function interactions (the Lieb-Liniger 
model \cite{LiebLiniger}) and its limit when the repulsion strength goes to infinity 
(the Tonks-Girardeau limit \cite{Tonks,Girardeau}). The latter is equivalent to free fermions. 
These systems only differ in the specific form of 
the phase shifts, relevant away from the transition. Eigenstates and thermodynamics 
\cite{Yang:1968rm} 
of these systems can be obtained from Bethe ans\"atze. We show that our data match well with the corresponding 
analytical nonrelativistic or relativistic predictions. Our simulations are performed at a lattice coupling in the continuum 
scaling regime, and we believe that the continuum limit to be performed does not reveal new qualitative features.

Our study demonstrates that  dual lattice simulations are capable of describing a system all the way from the 
Lagrangian in terms of ultraviolet fields to the infrared physics in terms of interacting particles (at nonzero densities 
induced by $\mu$). Such a transformation is the essence of Wilson's renormalization group, which is probably very hard 
to tackle analytically, but specific questions can be answered by the lattice, now that we have reliable lattice simulations 
at hand. To better understand the structure of dual partition functions and observables should also be of help in this 
program.

We briefly mention at the end that, although this program is mostly 
inspired by the attempts to understand the nonzero density structure of QCD, the potential benefit of studying nonlinear 
sigma models goes past this problem. Firstly, they are interesting in itself and appear as effective models of (anti-) 
ferromagnetic systems. Secondly, the study of properties of these systems is interesting in the context of continuum 
quantum field theories themselves. On the one hand, we have recently shown for the example of the O(3) nonlinear 
sigma model that lattice dualities may provide a physical connection with the low energy excitations of the theory. On the 
other hand, the nonlinear sigma models play a crucial role in the development of the continuum definition of quantum 
field theories via the resurgence program (see \cite{Dunne:2015eaa} and references therein). To date, resurgent 
constructions were explicitly shown to work only in analytical tractable one-dimensional reductions of the O(N) 
\cite{Dunne:2015ywa} and CP(N) \cite{Dunne:2012ae} nonlinear sigma models as well as of the principle chiral model 
\cite{Cherman:2013yfa}. However, genuinely 1+1 dimensional nonzero density systems, akin to what we study here, also 
show similar resurgence structures \cite{Fateev:1994ai}. Numerical and physical understanding of these systems is 
therefore important for the fundamentals of quantum field theory as well.

\section{Definition of the model and its lattice discretization}

\subsection{Continuum formulation and observables}

The O(3) model is conventionally written in terms of normalized vectors  $\vec{r}(x)=(r_1(x),r_2(x), r_3(x))$ with $\vec{r}(x)^2=1\:\forall x$ (also called `spins' or `O(3) rotors') and the continuum action reads  \cite{Hasenfratz:1990zz},
\begin{align}
 S[\vec{r}\,] \; 
 = \; \frac{1}{g^2} \int \! d^2x 
 &\Big[ \, \frac{1}{2}\, \big (\partial_\nu r_a \big)^2 + 
 i\,\mu\,\big(r_1\partial_2 r_2-r_2 \partial_2 r_1\big)
 \notag\\
 &+\frac{\mu^2}{2}\,r_3^2-\frac{\mu^2}{2}  \, \Big]~,
\label{eq:def_action_cont} 
\end{align}
where the coupling constant $g$ is dimensionless in 1+1 dimensions. We have already coupled a chemical 
potential $\mu$ to one of the O(2) subgroups, which excites the 3-component 
of the angular momentum. Repeated indices are 
summed over (\(\nu=1,2\) and \(a=1,2,3\)) and arguments $x$ have been dropped. 
At nonzero temperature $T$ the Euclidean 
time $x_2$ is periodic with period $1/T$. In such a bosonic theory, $\mu$ also enters quadratically tending to suppress 
the perpendicular component $r_3(x)$.

Our main thermodynamic observables will be the expectation values of the charge $Q$, its density $n$ 
and its susceptibility $\chi_n$: 
\begin{equation}
 Q
 =T\,\frac{\partial \ln Z}{\partial \mu}~,\qquad
 n
 =\frac{Q}{L}~,\qquad
 \chi_n 
 = \frac{\partial n}{\partial \mu}~,
\end{equation}
where $Z$ is the grand canonical partition function (see \cite{BruGaKloSu_1} or Eq.~\eqref{eq_dualZ} below). Eventually, 
all dimensionful quantities like $\mu, T, L, n$ etc.\ will be given in units of the mass $m$, e.g.,
\begin{equation}
 \frac{n}{m}
 =\frac{T/m}{Lm}\,
 \frac{\partial \ln Z}{\partial (\mu/m)}~,
\end{equation}
whereas $\chi_n$ is already dimensionless.

We will also explore the spin stiffness for which one imposes twisted spatial boundary conditions for a finite spatial length 
$L$. To implement these, we first introduce the O(2) polar angle $\phi$ by combining the first two components into a 
complex number 
\begin{equation}
r_1(x)+ir_2(x)=r_{12}(x)e^{i\phi(x)}~,
\end{equation}
and then replace the periodic boundary conditions in space by twisted ones,
\begin{equation}
 \phi(x_1+L,x_2)=\phi(x_1,x_2)+\ph~.
 \label{eq_bc_phi}
\end{equation}
If the twist costs free energy, $F = -T \ln Z$, at leading order the dependence of $F$ on $\varphi$ 
is quadratic in $\varphi$, and we define the \textit{spin stiffness} (also called superfluid density) by 
\begin{equation}
 \sti
 =L\,\left.\frac{\partial^2 F}{\partial\ph^2}\right|_{\ph=0}
 =-L\,T\,\frac{1}{Z}\,\left.\frac{\partial^2 Z}{\partial\ph^2}\right|_{\ph=0}~,
 \label{eq_def_stiff}
\end{equation}
where we have used that $Z$ is an even function of $\ph$.
Physically it is clear that $\sti$ depends on whether the regions $x_1$ and $x_1+L$ are correlated, i.e., whether the 
system is in a (spatially) ordered state.

The spin stiffness can be computed and related to vortices in the lattice O(2) model without chemical potential\footnote{The O(2) lattice action is $-J\sum_{x,\nu}\cos(\phi(x+\hat{\nu})-\phi(x))$, which is nothing but 
\eqref{eq_lattice_action} with $r_3=0$, $r_{12}=1$ at $\mu=0$.} \cite{Mudry}. For large lattice coupling, the vortices 
arrange in pairs and the spin correlator decays algebraically, which is the behavior closest to an ordered state in two 
dimensions (as the Mermin-Wagner theorem forbids the spontaneous breaking of the continuous symmetry). As a 
consequence, the spin stiffness $\sti$ will be nonzero. At small lattice coupling, the vortices condense and make the 
correlator decay exponentially. In this regime the spin stiffness $\sti$ will vanish if $L$  is larger than the spatial 
correlation length $\xi$. This is why the spin stiffness can be used to detect BKT transitions characterised by 
the change of the correlator decay and to measure the spatial correlation length. 

Actually, the dimensionful combination $\ph/L$ may be viewed as an imaginary chemical potential in the spatial direction,
and therefore the stiffness is known to measure spatial winding numbers \cite{Pollock:1987}. As this is best seen using dual variables, we give the corresponding formula in the next section. 

\subsection{Lattice formulation and the dual representation}
\label{sec_lattice_action_etc}
 
The lattice action discretizing \eqref{eq:def_action_cont} reads,\begin{align}
 S[\vec{r}\,] 
 =& - J\sum_{\substack{x\in\Lambda\\ \nu=1,2}}
 \Big[ r_{3}(x)\, r_{3}(x+\hat{\nu})+  \frac{1}{2} \,  r_{12}(x) \,  r_{12}(x+\hat{\nu})
 \notag\\
 &\times\big\{e^{-i(\phi(x)-\phi(x+\hat{\nu}))-\mu\,\delta_{\nu,2}} + 
 c.c.|_{\mu\to -\mu}\big\}\Big] \; .
 \label{eq_lattice_action}
\end{align}
As common in lattice field theory, the chemical potential $\mu$ introduces exponential factors for the forward and 
backward temporal hopping terms. For $\mu = 0$ these terms are related by complex conjugation $c.c.$, but 
when $\mu$ has a nonzero real part we face a complex action problem.
The parameter $J$ is the lattice coupling constant (dimensionless and positive) and the first sum runs over the $V \equiv 
N_s\times N_t$ sites of a two-dimensional lattice \(\Lambda\) with periodic boundary conditions.
Again $\nu=1,2$ and $\hat{\nu}$ denotes the corresponding unit vector in direction $\nu$. Throughout this paper we set 
the lattice spacing $a$ to $a = 1$, implying that $T=1/N_t$ and $L=N_s$.

The continuum limit for the lattice model is reached via $J\rightarrow \infty$. The mass gap of the 
system can be expressed in the bare coupling $J$ and a UV cut-off as $m^2=\Lambda_{UV}^2\exp(-4\pi J)$. On the 
lattice the cut-off is proportional to the inverse lattice spacing, $\Lambda_{UV}=C/a$, and to two loops the mass gap 
reads\cite{twoloopnum1,twoloopnum2},
\begin{align}
 am=C(1+2\pi J)\exp(-2\pi J) \quad\text{for } J\to\infty~.
 \label{eq_two_loop_massgap}
\end{align}
The partition sum is defined as the lattice path integral
$Z=\int \mathcal{D}[\vec{r}\,] \,e^{-S[\vec{r}]}$, where the measure $\mathcal{D}[\vec{r}\,]$ is the product over the 
O(3)-invariant measures
for $\vec{r}(x)$  on all lattice sites. The particle density and susceptibility are defined as $\mu$-derivatives of $Z$ as in 
the continuum,
\begin{equation}
 n=\frac{1}{N_s N_t}\,
 \frac{\partial \ln Z}{\partial \mu}~,\qquad
 \chi_n 
 = \frac{\partial n}{\partial \mu}~.
 \label{eq_density_susc_first}
\end{equation}
As a check we will also use the expectation value of the action density at $\mu=0,$\footnote{This is obtained 
from $e=\langle E\rangle/N_sN_t$, where 
$E=\sum_{x,\nu}(\nabla_\nu\vec{r})^2$ and $(\nabla_\nu f)(x)=f(x+\hat{\nu})-f(x) $ \cite{BergLuscher}. From the 
normalization of $\vec{r}$ it follows that $e=4-2\langle \vec{r}(x)\vec{r}(x+\hat{\nu})\rangle/
N_sN_t$ and thus \eqref{eq_energy_first}.}
\begin{equation}
 e
 =4-\frac{2}{N_s N_t}\,
 \frac{\partial \ln Z}{\partial J}~,
 \label{eq_energy_first}
\end{equation}
whose strong and weak coupling expansions are \cite{BergLuscher}, 
\begin{equation}
 e =\begin{cases}
 4-4y-8y^3-\frac{48}{5}y^5+\ldots,%\:\text{where }y=\coth J-\frac{1}{J}
 & \text{for small } J,\\
 \frac{2}{J}+\frac{1}{4J^2}+\frac{0.156}{J^3}+\ldots & \text{for large } J.
 \end{cases}
 \label{eq_energy_expansions}
\end{equation}
where $y=\coth J-\frac{1}{J}$.

In \cite{BruGaKloSu_1} we have introduced the following (exact) representation of the partition function in terms of integer 
dual variables $m_{x,\nu}\in\mathbb{Z}$ and $k_{x,\nu},\overline{m}_{x,\nu}\in \mathbb{N}_0$,
\begin{widetext}\begin{align}
Z= \label{eq_dualZ}
& \sum_{\{m,\overline{m},k\}} \left(\prod_{x,\nu} \frac{J^{k_{x,\nu}}}{k_{x,\nu}!} \frac{(J/2)^{|m_{x,\nu}|+
2\,\overline{m}_{x,\nu}}}{(|m_{x,\nu}|+\overline{m}_{x,\nu})!~\overline{m}_{x,\nu}!} \right) e^{\,\mu\sum_x m_{x,2}}   \\
& \times ~ \prod_x \mathcal{I}\left(\sum_\nu(k_{x,\nu}+k_{x-\hat{\nu},\nu}),1+\sum_\nu\Big[m_{x,\nu}+
m_{x-\hat{\nu}}+2\,(\overline{m}_{x,\nu}+\overline{m}_{x-\hat{\nu}})\Big]\right) \nonumber \\
& \times ~ \prod_x \delta\left(\sum_\nu[m_{x,\nu}-m_{x-\hat{\nu},\nu}]\right) ~ E\left(\sum_\nu[k_{x,\nu}+k_{x-\hat{\nu},\nu}]\right)~, \nonumber
\end{align}\end{widetext}
where the Kronecker delta, the evenness function $E$ and a function $\mathcal{I}$ (related to the beta function) have been used:
\begin{align}
 \delta(n) 
 &=
\begin{cases}
1 &n=0 \\
0 &\text{else}
\end{cases}
~ ,\quad 
E(n) =
\begin{cases}
1 &n~\text{even} \\
0 &n~\text{odd}
\end{cases}~,\notag\\% \quad \text{and}\\
 \mathcal{I}(a,b)
 &=\frac{\Gamma\left(\frac{a+1}{2}\right)\,\Gamma\left(\frac{b+1}{2}\right)}{\Gamma\left(\frac{a+b+2}
{2}\right)}~.
\end{align}
Note that the current $m_{x,\nu}$ is conserved since the Kronecker delta of $\sum_\nu [m_{x,\nu}-m_{x-\hat{\nu},\nu}]
\equiv(\nabla m)_{x}$ corresponds to the discrete version of the vanishing divergence condition (at each site $x$). The 
chemical potential couples to the corresponding sum over the temporal components of $m_{x,\nu}$. 
This sum can be rewritten 
using the conservation as $\sum_x m_{x,2}=N_t\sum_{x_1}m_{x,2}=1/T\cdot w[m]$, where $w[m]$ is the total winding 
number of the $m$-loops reflected in the net $m$-flux through every time slice. Thus, the chemical potential appears 
through weights $\exp(\mu\cdot\text{integer}/T)$. 
This is also the main advantage of dual representations of systems with respect to chemical 
potentials: if the dual partition function has no sign problem at vanishing $\mu$ -- which holds for our system -- $\mu$ 
does not introduce a sign problem either.

In the dual representation, the observables take the form
\begin{align}
 n
 &=\frac{1}{N_s N_t}\big\langle \,\sum_x m_{x,2}\big\rangle
 =\frac{1}{N_s}\big\langle w[m]\big\rangle~,
 \label{eq_observables_dual_zero_densonly}\\
 \chi_n
 &=\frac{N_t}{N_s}\big(\langle w[m]^2\rangle-\langle w[m]\rangle^2\big)~,\label{eq_observables_dual_zero_chionly}\\
 e
 &=4-\frac{2}{N_s N_t}\,
\big\langle \,\sum_{x,\nu}\big[ \, k_{x,\nu} + |m|_{x,\nu} + 2 \, \overline{m}_{x,\nu} \, \big]\big\rangle~,
\label{eq_observables_dual_one}
\end{align}
where here $\langle O\rangle$ is the expectation value of $O$ in the dual representation  
obtained by inserting the expression $O$ into the sum in 
Eq.~\eqref{eq_dualZ} and dividing by $Z$.

We will also measure the space-time average of the third dual variable, 
\begin{equation}
K \; =\; \frac{1}{N_sN_t}\,\big\langle \,\sum_{x,\nu}\, k_{x,\nu}\big\rangle~,
\label{eq_thirddual}
\end{equation}
as a measure for the anisotropy of the system. For its interpretation we sketch 
how the dual representation is obtained for the third variable, using that 
$Z\propto\int \mathcal{D}[\vec{r}\,] \prod_{x,\nu}\sum_{k_{x,\nu}=0}^\infty \big[J\,r_3(x)r_3(x+
\hat{\nu})\big]^{k_{x,\nu}}/k_{x,\nu}!$. Evaluating (\ref{eq_thirddual}) inserts another $k_{x,\nu}$ 
into the dual partition sum and reduces the 
argument of the factorial by one, which can be compensated by a shift of the summation variable giving a factor of $J$ 
and the hopping term. It follows that $\langle k_{x,\nu}\rangle=J\,\langle r_3(x)r_3(x+\hat{\nu})\rangle$, 
where the latter expectation value is in the conventional representation 
\eqref{eq_lattice_action}. Hence $K$ measures the 
amount of hopping in the direction perpendicular to the $x-y$ plane where we excite O(2) angular momentum. Thus, we 
expect $K$ to be small in the anisotropic phase at large $\mu$.

At vanishing $\mu$ the theory enjoys the full O(3) symmetry and the amount of hopping must be the same for all 
components. In the energy density  \eqref{eq_observables_dual_one} above this can be seen by virtue of the fact that 
$|m|_{x,\nu} + 2 \, \overline{m}_{x,\nu}$ is the sum of two dual variables of the same nature as $k_{x,\nu}$;
see Eq.\ (12) of \cite{BruGaKloSu_1}. Therefore, $e=4-6K/J$ should hold at $\mu=0$.

In the same way the chemical potential couples to the integrated temporal component of the conserved O(2) current, the 
twist-induced imaginary spatial chemical potential $i\ph/L$ couples to the integrated spatial component. The partition 
function in the presence of the twist thus has an additional factor $\exp(i\ph/N_s\cdot \sum_x m_{x,1})=\exp(i\ph\, 
w_s[m])$, where $w_s[m]$ is the total spatial winding number of $m$ in each configuration. In the definition of the spin 
stiffness $\sti$, Eq.~\eqref{eq_def_stiff}, the second derivative with respect to $\ph$ brings down $-w_s[m]^2$ and 
setting $\ph$ to zero afterwards results in the expectation value one obtains
\begin{align}
 \sti=\frac{N_s}{N_t}\, \langle w_s[m]^2\rangle
 =LT\, \langle w_s[m]^2\rangle\; ,
\end{align}
in the dual repesentation without twist (similar to $\chi_n$ in Eq.~\eqref{eq_observables_dual_zero_chionly}).

\section{Numerical simulation, tests and basic analysis}

In this section we collect several more technical aspects of this paper. We briefly discuss our simulation strategy for 
the dual formulation and evaluate the correctness of its results by comparing them to perturbative strong and weak 
coupling calculations. Furthermore we present numerical results for the phase diagram at finite volume and temperature
as well as a finite volume scaling analysis which indicates that at nonzero temperature all transitions are smooth
crossovers.

\begin{figure}[!b]
\includegraphics[width=\linewidth,type=pdf,ext=.pdf,read=.pdf]{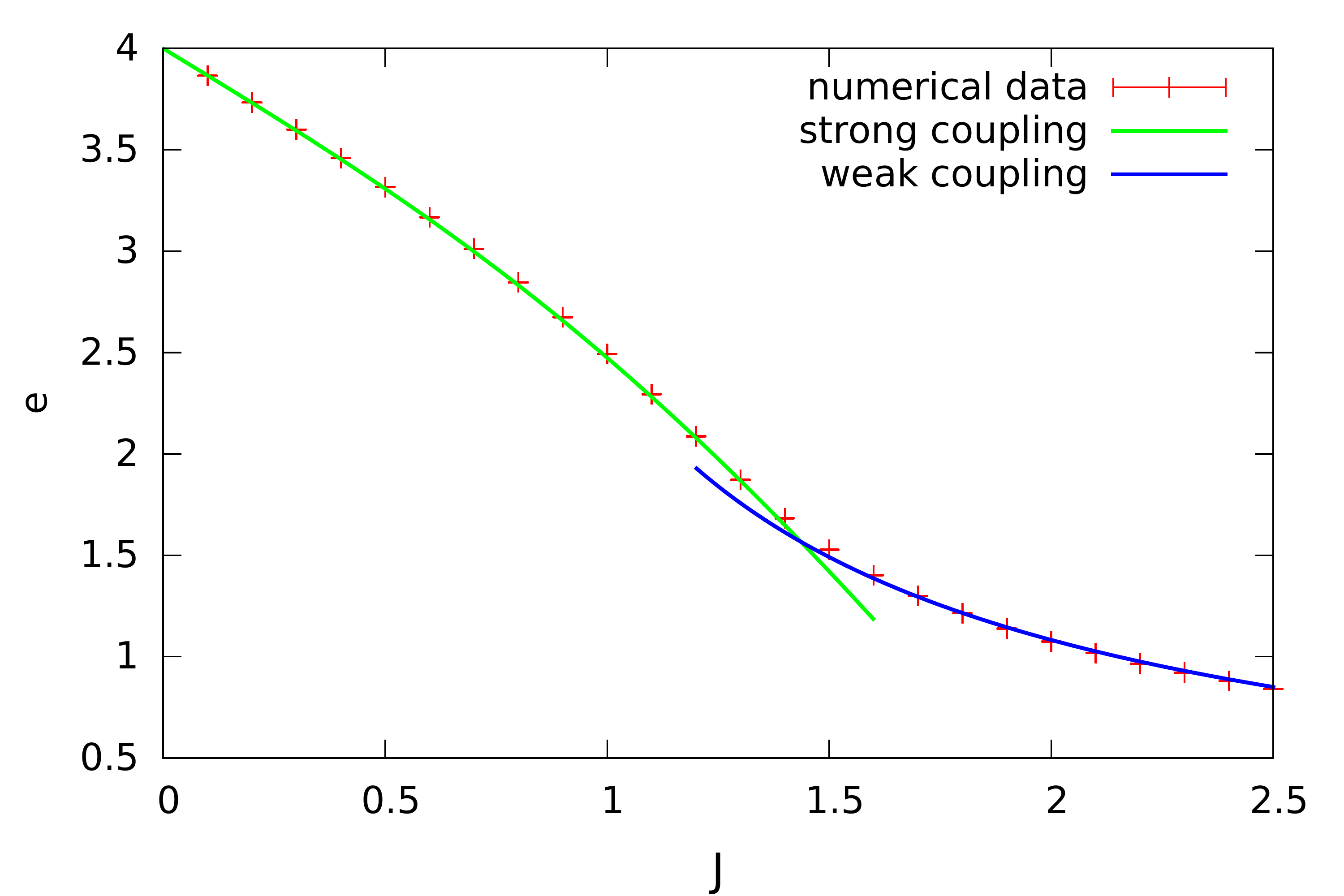}
\caption{Energy density versus the coupling at \(\mu=0\). The analytical results for the strong and weak coupling 
expansions from Eq.~\eqref{eq_energy_expansions} agree very well with the numerical results (having very small error 
bars) obtained from simulating the dual ensemble on a  \(10\times10\) lattice and using 
Eq.~\eqref{eq_observables_dual_one}.}
\label{fig_energy}
\end{figure}

\subsection{Dual simulation and tests} 

In this subsection we briefly discuss Monte Carlo simulation strategies for the dual representation Eq.~(\ref{eq_dualZ})
and present tests for its correctness. Obviously each term in the partition sum (\ref{eq_dualZ}) is real and positive 
and a probability interpretation of the weights of the dual configurations is possible. 
The remaining challenge of a dual Monte Carlo simulation is to generate only those configurations 
that obey all the constraints. For the unconstrained dual variables $\overline{m}_{x,\nu}$ conventional local Metropolis 
updates are sufficient. For the constrained variables $m_{x,\nu}$ and $k_{x,\nu}$ the constraints enforce closed loops 
and loops where flux is conserved modulo 2, respectively. In both cases one can generate new admissible configurations 
by changing the variables along an arbitrarily chosen closed loop which guarantees that the constraints remain intact. 
This loop along which one updates the dual variables can for example be grown 
in steps using local random choices and corresponding Metropolis decisions until it closes, which is the well known worm 
strategy \cite{worm} -- this is the update used here. After equilibration, we typically use $10^5$ to $10^6$ measurements 
on configurations separated by ${\mathcal O}(20)$ sweeps. The statistical errors we show are 
determined with the jackknife method taking autocorrelation times into account.

\begin{figure}[!t]
\includegraphics[width=\linewidth,type=pdf,ext=.pdf,read=.pdf]{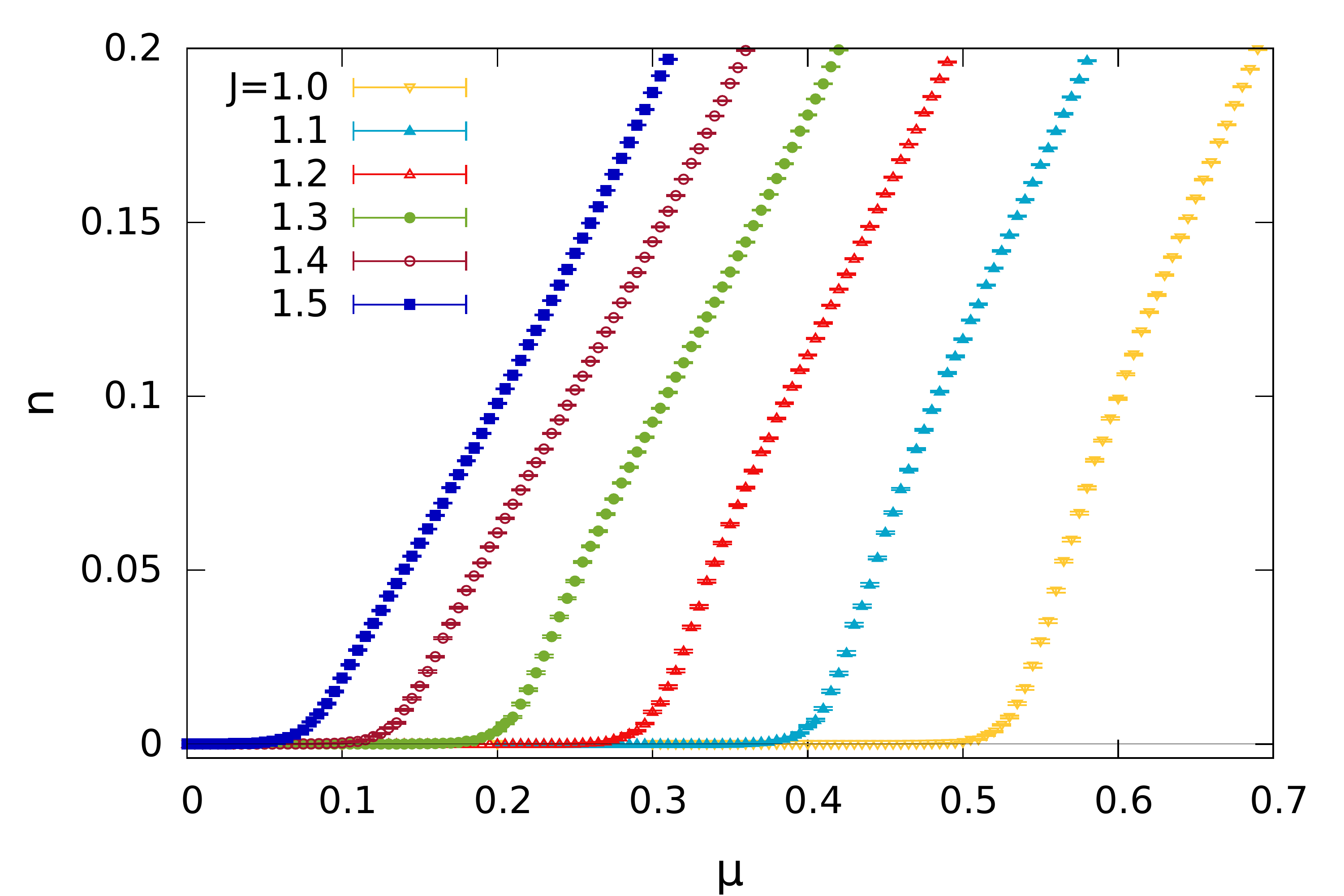}

\includegraphics[width=\linewidth,type=pdf,ext=.pdf,read=.pdf] {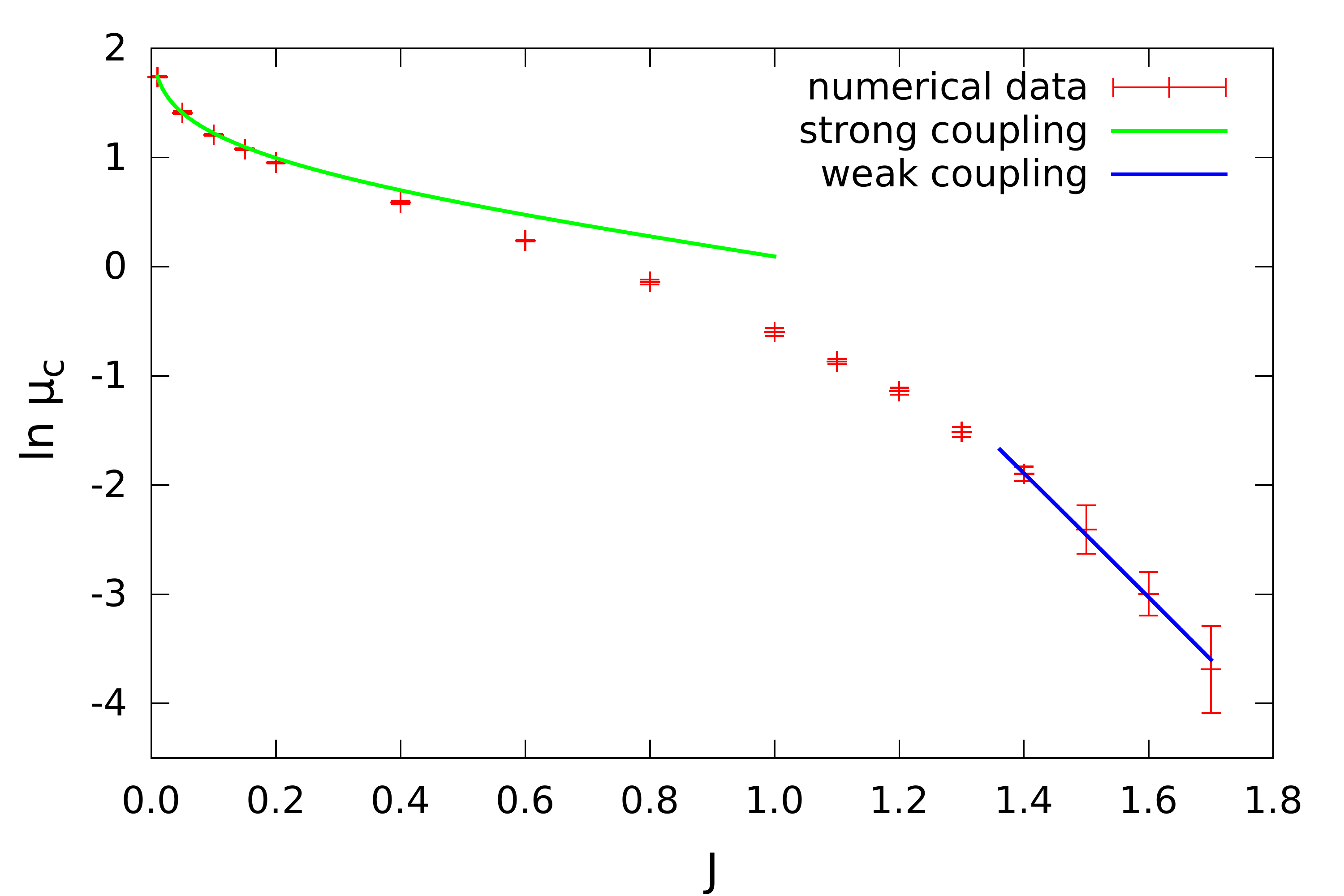}
\caption{Top: Expectation value of the particle number density versus the chemical potential $\mu$, both in lattice units, 
at different couplings $J$ for lattice size $90 \times 90$. Bottom: Logarithm of the corresponding critical chemical 
potentials as a function of the coupling. We compare our data to the strong coupling result $\ln(3/J)$ and the  
weak coupling expansion of the mass 
gap and Eq.~\eqref{eq_two_loop_massgap}, having obtained  $C=102(2)$ from a fit of the $J\geq 1.4$ data.}
 \label{fig_part_vs_mu_first}
\end{figure}

For a first test we computed the energy density at vanishing chemical potential, 
cf.~Eq.~\eqref{eq_observables_dual_one}. In Fig.~\ref{fig_energy} we show the results for 
$e$ at $\mu = 0$ as a function of the coupling $J$ and compare to the weak and strong coupling expansions 
from Eq.~\eqref{eq_energy_expansions}. We find excellent agreement in the 
corresponding domains of $J$. This demonstrates that the mapping to the dual variables and the implementation of the 
dual Monte Carlo simulation are correct.

We now switch to the situation where the dual approach is really essential, i.e., the simulations at nonzero chemical 
potential.  In the top panel of Fig.~\ref{fig_part_vs_mu_first} we plot the particle density $n$ measured via 
Eq.~\eqref{eq_observables_dual_zero_densonly} as a function of $\mu$. As a general phenomenon at low temperatures a net 
density is induced into the system only after $\mu$ has reached a \textit{threshold}, the mass of the lightest 
particle with charge coupling to $\mu$. The critical values of $\mu$ 
visible in that plot thus depend on the coupling $J$ just like the mass. This is shown in the bottom panel of 
Fig.~\ref{fig_part_vs_mu_first}, again with strong and weak coupling expansions. The former, $\mu=\ln(3/J)$ for $J\to 0$, 
is worked out in App.~\ref{app_strong_coupling_mus}, whereas the latter follows from the two-loop mass gap formula, 
Eq.~\eqref{eq_two_loop_massgap}, in which we obtain the constant $C$ related to the UV cut-off
by a one-parameter fit.  The agreement is again very good, which is seen also in a comparison to the literature 
(Fig.~10 of \cite{twoloopnum2}).
Note also that the continuum scaling sets in at $J\simeq 1.4$, as for the energy density in Fig.~\ref{fig_energy}.

Most of our lattice data were taken at $J=1.3$, where $am \sim 0.22$.
This means a restriction to $\mu\ll 5m$, otherwise $a\mu$ becomes comparable to 1 and strong 
discretization effects set in.

\subsection{Finite lattice phase diagram}
\label{sec_lattice_diagram}

\begin{figure}[b]
\includegraphics[width=\linewidth,type=pdf,ext=.pdf,read=.pdf]
{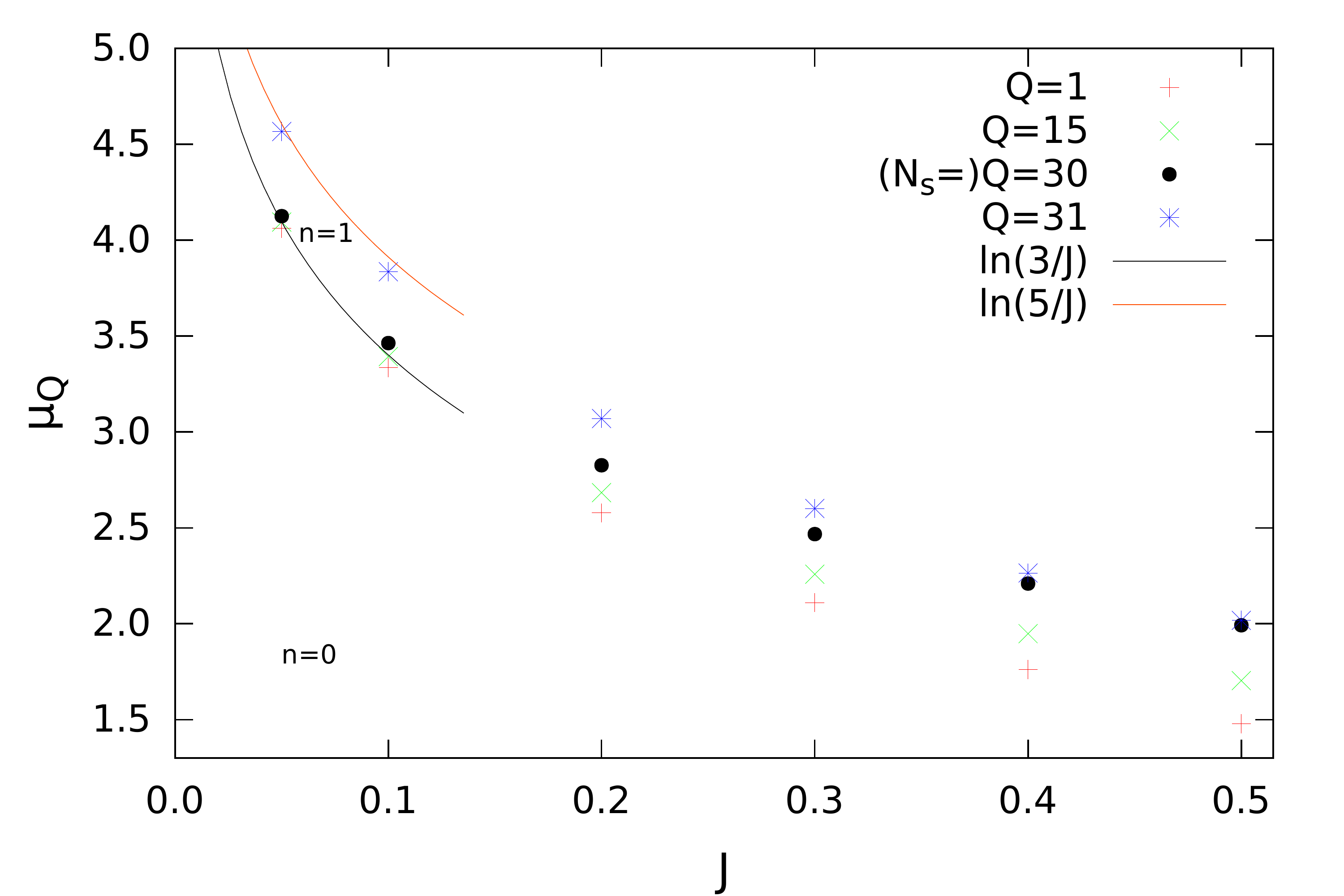}
\caption{Phase transition in the $J$-$\mu$ plane for $N_s=30$ and $N_t=1000$. At large $J$, i.e., towards the 
continuum limit, the critical chemical potentials with increasing index (from red to green to black to blue) induce individual 
transitions between integer charges as utilized in \cite{BruGaKloSu_2}. At small $J$ the first $N_s$ of them join to a 
single transition, $\mu_1=\ldots=\mu_{Ns}$,  increasing the particle \textit{density} $n$ in lattice units from 0 to 1 (see 
text). The next critical chemical potential $\mu_{N_s+1}$ (blue) is separated from those, and above it a density $n=2$ 
will be induced. At large $J$, on the other hand, the change from $\mu_{Ns}$ to $\mu_{N_s+1}$ plays no particular role. 
The strong coupling predictions for the transitions $0\to 1$ and $1\to 2$ are included as solid curves.}
\label{fig_many_mucs}
\end{figure}

Before we come to discussing continuum results, we first consider nonzero lattice spacing and finite volume,
i.e., we study the system at fixed lattice size 
without a final continuum limit $J\to\infty$. We determine the $J$-$\mu$ phase 
diagram at low temperature, using the onset of non-zero particle density as a function of $\mu$ for determining
the phase boundaries. Fig.~\ref{fig_many_mucs} shows this phase diagram for several critical chemical potentials. 
Note that the number of temporal sites is fixed, so the temperature in mass units $T/m=1/(N_t am)$ varies with the coupling according to the mass gap formula $am(J)$, e.g.,
Eq.~\eqref{eq_two_loop_massgap}.

In \cite{BruGaKloSu_2} we have shown that close to the continuum limit at small temperatures and volumes, the 
particles are induced into the system one by one displaying integer plateaus in the particle number itself (not its density). 
These transitions show up for large $J$ and some examples are plotted in Fig.~\ref{fig_many_mucs}. The locations of 
these critical $\mu$'s are governed by the particle interaction phase shifts and thus contain interesting physics, as 
explained in detail in \cite{BruGaKloSu_2}.

In the strong coupling regime at small $J$ the situation is different. In the dual representation of the partition function, 
Eq.~\eqref{eq_dualZ}, every dual variable is suppressed by the corresponding power of $J$.
Still, the temporal components of the flux variable $m$ are promoted by factors of $e^\mu$, that eventually overcome the 
factors of $J/2$. This mechanism acts locally on every spatial site, which means that if it is preferable to have $Q$ 
particles at some site ($m_{x,2}=Q$), then this immediately applies to all sites. In fact, the 
superposition of $Q$ fluxes does not cost any action at strong coupling. Therefore, in this regime the particle number 
\textit{density} changes by one (in lattice units). The corresponding critical values of $\mu$ 
depend on $J$ logarithmically, and their 
values are computed in Appendix \ref{app_strong_coupling_mus}. Fig.~\ref{fig_many_mucs} shows good agreement 
with these curves for small $J$ and illustrates 
how intermediate couplings interpolate the transitions between these regimes. In 
particular all critical values of $\mu$ are on equal footing towards the continuum 
(large $J$), whereas in the strong coupling 
regime (small $J$) multiples of $N_s$ bunch, such that 
regimes with fixed lattice density open up.

\begin{figure}[b]
\includegraphics[width=\linewidth,type=pdf,ext=.pdf,read=.pdf]
{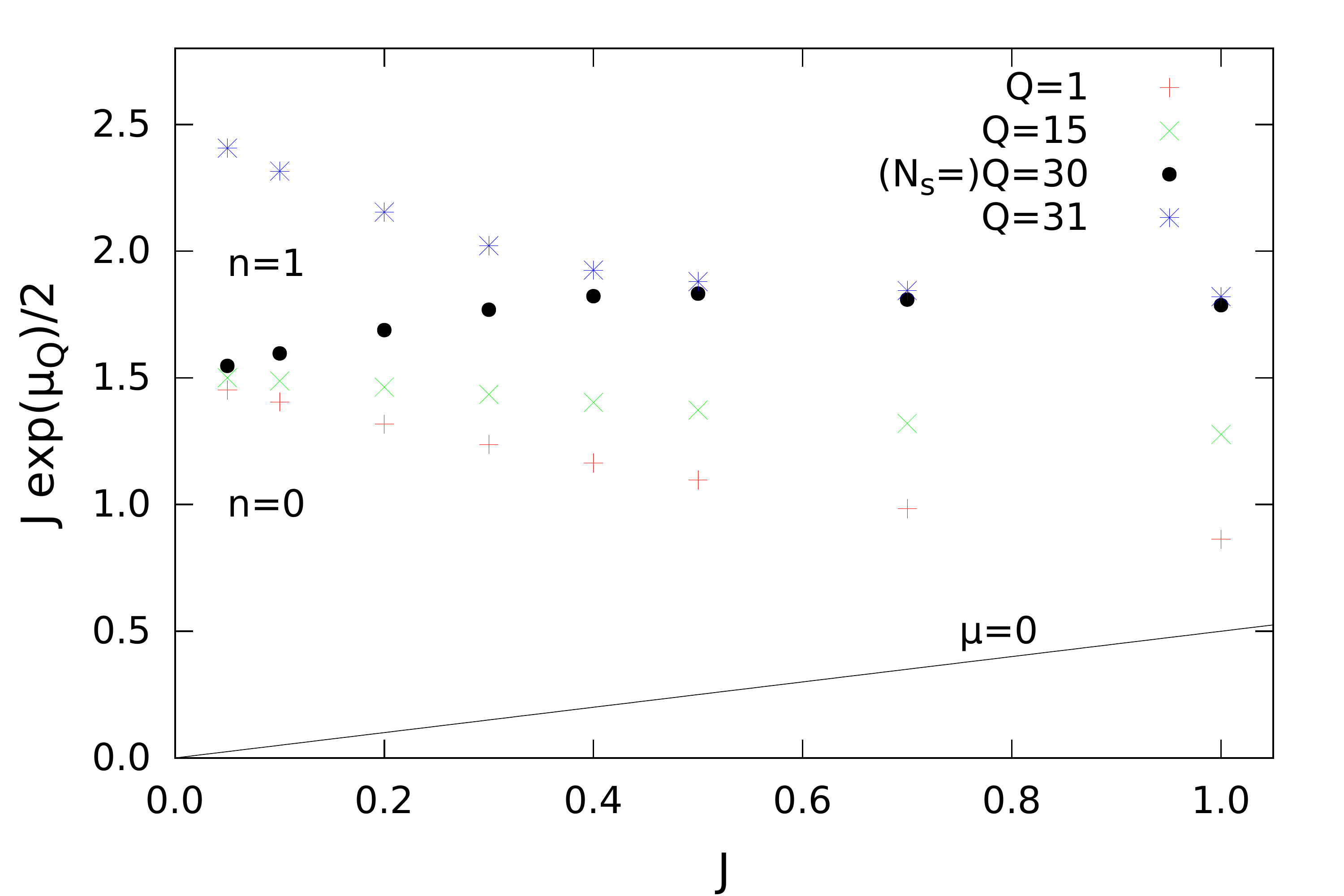}
\caption{Phase transition as in Fig.~\ref{fig_many_mucs}, but with a modified quantity on the $y$-axis (and some more data points for higher $J$'s), see text.}
\label{fig_many_mucs_o2}
\end{figure}

In Fig.~\ref{fig_many_mucs_o2} we show the lattice phase diagram with the quantity $J/2\cdot \exp(\mu_Q)$ on the $y$-axis. In the strong coupling limit $N_s$ critical $\mu$'s now meet at half integers 3/2, 5/2 etc., which agrees with the derivation in Appendix  \ref{app_strong_coupling_mus} again. The corresponding diagram for the O(2) model is shown in Fig.~3 of Ref.~\cite{o2} (for this model the curves meet at integer values). 

\begin{figure}
 \includegraphics[width=\linewidth,type=pdf,ext=.pdf,read=.pdf]{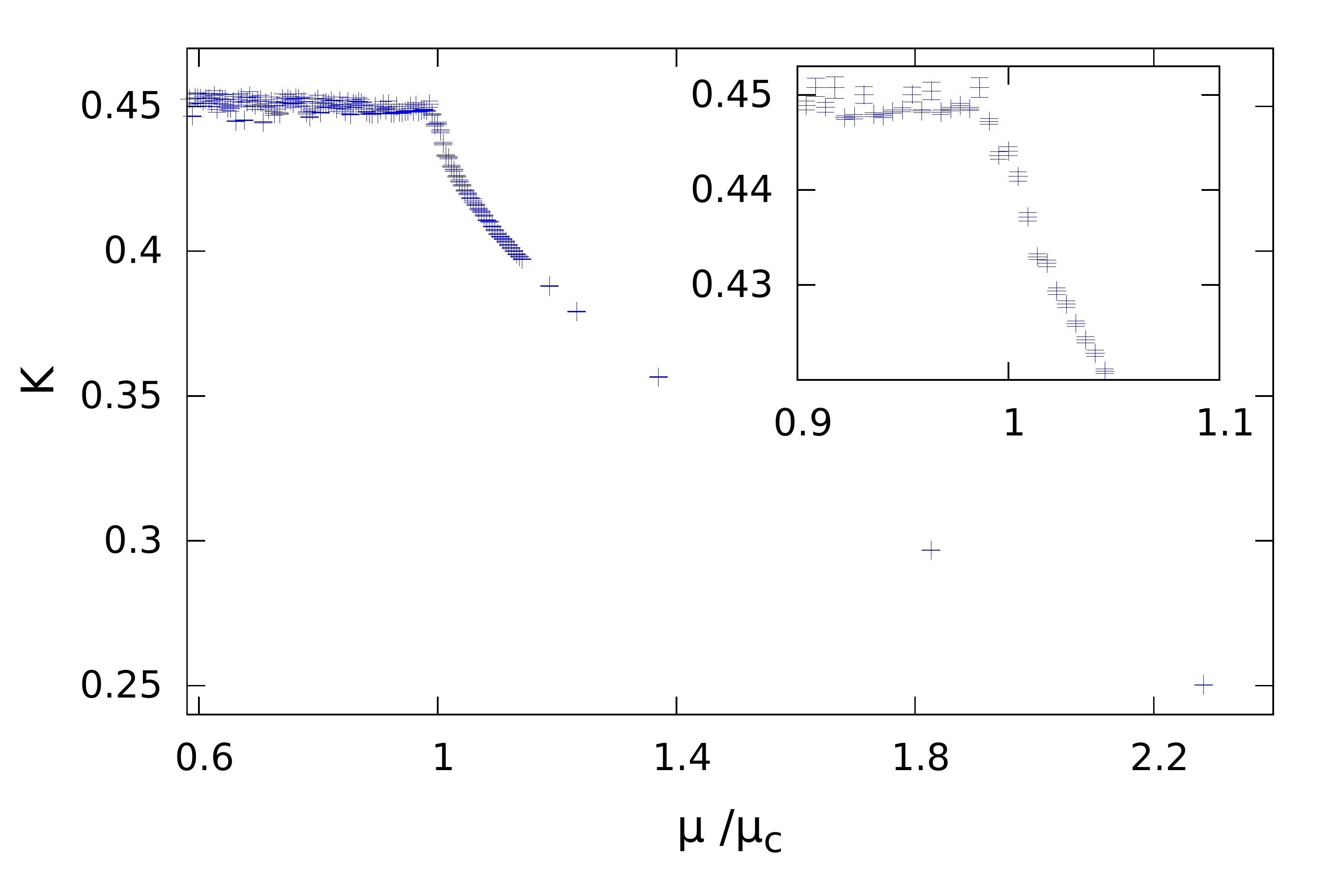}
\caption{Expectation value $K$ for the third dual variable (for its definition and interpretation see 
Eq.~\eqref{eq_thirddual} and below) also displaying a transition at $\mu=m$ ($J=1.3, N_s=200, N_t=1000$). 
The value at  small $\mu$ is the isotropic one related to the energy density by $K=J(4-e)/6$.}
\label{fig_thirddual}
\end{figure}

Let us conclude this subsection with a quick look at the third dual variable $k_{x,\nu}$, or more specifically at its
sum $K$ as defined in Eq.~\eqref{eq_thirddual}. For $\mu$ larger than the threshold the O(3) symmetry 
is explicitly broken to O(2) by the presence of the O(2) charge, which is expected to be manifest 
in a decrease of $K$ (see the discussion below Eq.~\eqref{eq_thirddual}). 
In Fig.~\ref{fig_thirddual} we show $K$ as a function of $\mu$ (normalized by $\mu_c$) and indeed 
find the onset of a drop
at the critical chemical potential, which confirms the picture that the system tends to become more and more planar 
as $\mu$ increases.

\subsection{Crossover at nonzero temperature}
\label{section_finite_volume}

We conclude the first analysis of the lattice model by studying the nature of the transitions 
mapped out in the previous subsection. For this study we still keep the temperature $T$ fixed and 
perform the standard scaling analysis of order parameters with the spatial volume $L$ to analyze the 
nature of the transition.

\begin{figure}[!h]
 \includegraphics[width=\linewidth,type=pdf,ext=.pdf,read=.pdf,clip]{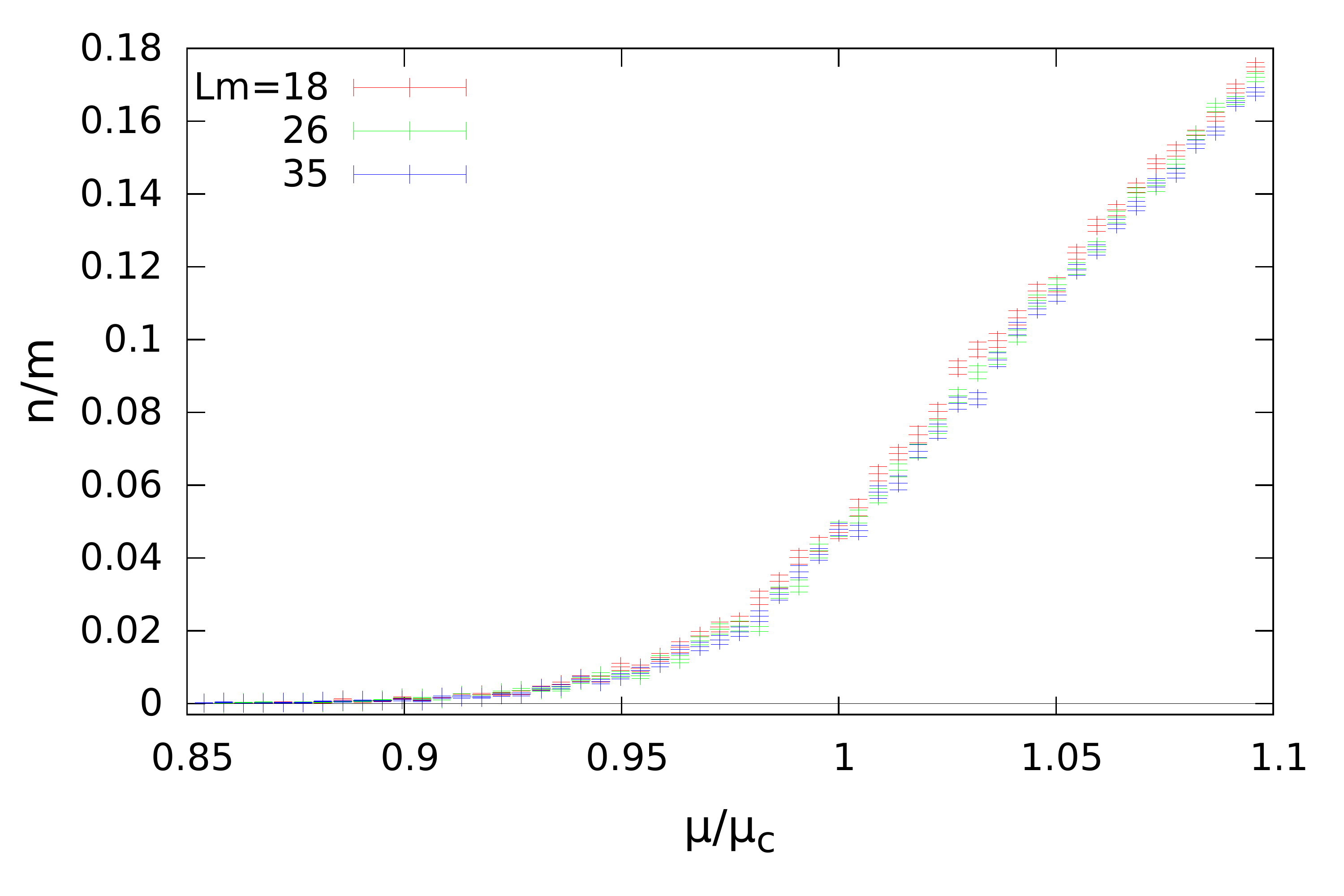}

 \includegraphics[width=\linewidth,type=pdf,ext=.pdf,read=.pdf,clip]{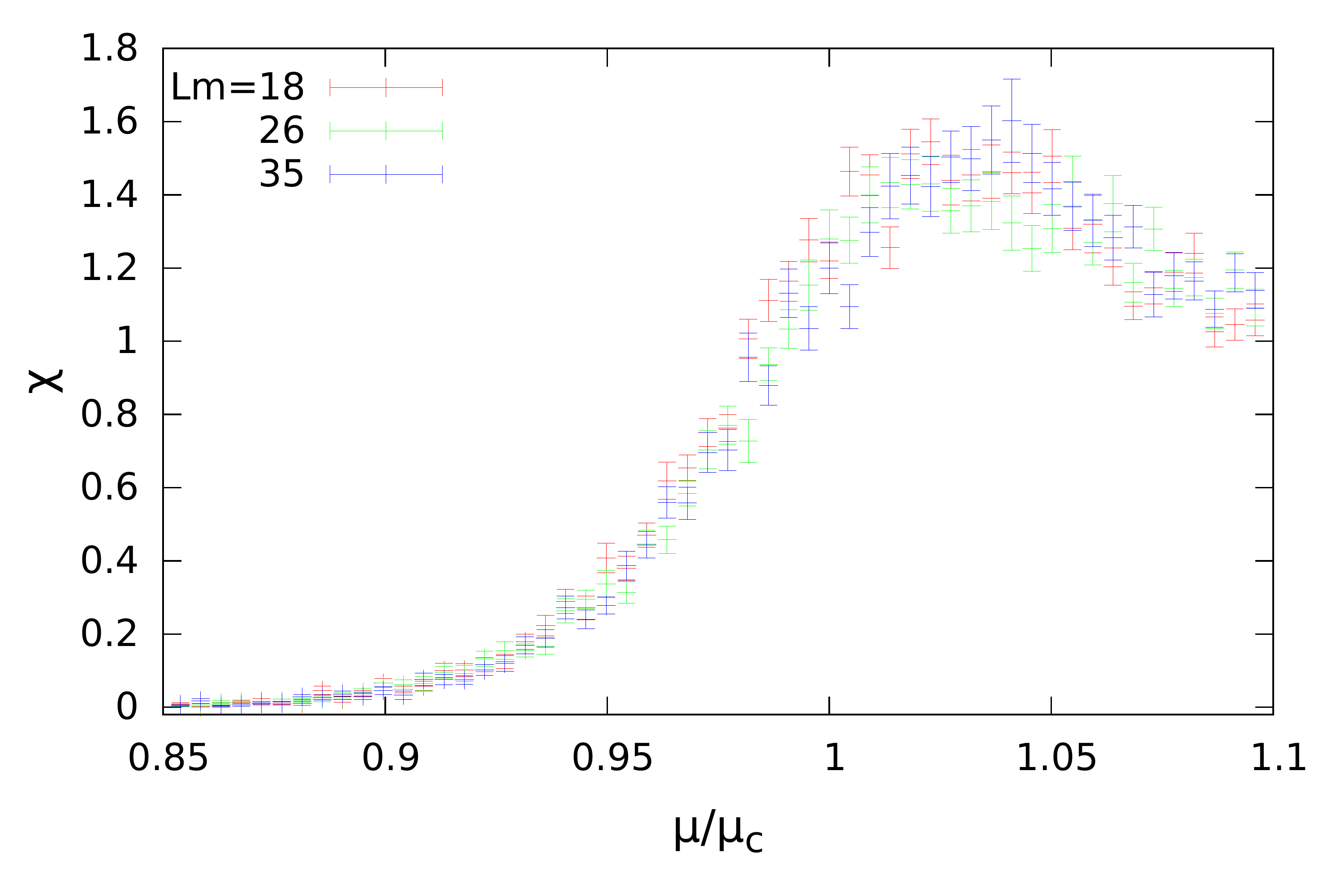}
\caption{Dependence of the particle number density and its susceptibility on the chemical potential at a fixed 
low temperature $T/m=0.023$ for three spatial sizes volumes (coupling $J=1.3$, lattices with $N_t=200$ and 
$N_s=80,\,120,\,160$).}
\label{fig_volumescaling}
\end{figure}

Fig.~\ref{fig_volumescaling} shows the particle number density (top panel) and its susceptibility (bottom panel)
at a fixed low temperature of 2\% of the mass. 
As a function of chemical potential $\mu$ the particle number $n$ is monotonically increasing, with a strong 
variation emerging only above $\mu_c$, while the susceptibility displays a maximum. 
Doubling the spatial size, the data still fall on top of each other for both observables.
Equivalent results were found for other nonzero temperatures and we conclude 
that for nonzero temperature the transition is smooth, i.e., it is a \textit{crossover}.

\section{Quantum phase transition}

We now explore the possibility of a quantum phase transition at zero temperature, i.e., we 
focus on the combined limit of zero temperature and infinite volume, i.e., 
$T \! =\!  0,\, L \!=\! \infty$ and analyze the transitions of the O(3) model
as a function of the chemical potential. This amounts to sending the extents of both, Euclidean time and space to infinity. 
Of course, with finite numerical simulations this can only be done as a limit, $T \rightarrow 0$ and $L^{-1} \rightarrow 0$. 
However, in general, the outcome depends on the particular choice of the trajectory towards that limit which 
one chooses in the $T$-$L^{-1}$ plane. 
In particular the spatial  and temporal correlation lengths can be different, as expressed by the dynamical critical 
exponent (see Sec.~\ref{sec_qpt_stiff} below). Thus we will discuss the behavior of the system for 
different "scaling trajectories", i.e., different paths 
leading to the limit $T \! =\!  0,\, L^{-1} \!=\! 0$.  For these scaling trajectories 
we will present and interpret our data for the particle density and the spin stiffness. 
For the interpretation we will partly rely on simple model systems showing characteristic features 
observed in our simulations.

\subsection{Scaling trajectories towards zero temperature and infinite volume and particle number results}

\begin{figure}[t]\centering
\begin{tikzpicture}
 \fill[black] (0,0) circle (0.15);
 \draw[<-, line width=1pt] (0,0.3) -- (0,4) -- (3.2,4);
 \draw[->, line width=1pt] (0,4.2) -- (0,5);
 \draw[<-, line width=1pt] (0.3,0) -- (4,0) -- (4,3.2);
 \draw[->, line width=1pt] (4.2,0) -- (5,0);
 \draw[<-, line width=1pt] (0.3,0.3) -- (3.4,3.4);
 \draw[<-, line width=1pt] (0.3,0.05) .. controls (2.2,0.25) and (3,0.5) ..  (3.85,3.2);
 \node (T) at (-0.4,4.8) {\Large $T$};
 \node (L) at (5.5,0) {\Large $L^{-1}$};
 \node (a0) at (1.9,4.3) {\Large $\alpha=0$};
 \node (a1) at (1.4,2.4) {\Large $\alpha=1$};
 \node (a2) at (2.35,1.43) {\Large $\alpha=2$};
 \node (ainf) at (4.9,1.9) {\Large $\alpha=\infty$};
\end{tikzpicture}
\caption{Illustration of the four scaling trajectories in the $T$-$L^{-1}$ plane 
towards the zero temperature and infinite volume limit
(black dot) which we use in this study (see discussion around \eqref{eq:scaling}). }
\label{fig_scaling_diagram}
\end{figure}
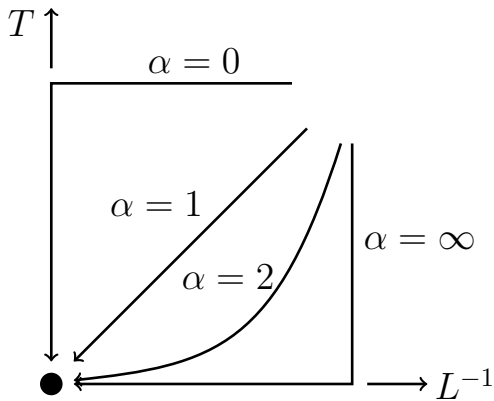

As already outlined, when considering the limits $T \rightarrow 0$ and $L^{-1} \rightarrow 0$, 
the behavior of the system will depend on the particular 
scaling trajectory one follows in the $T$-$L^{-1}$ plane towards the origin. 
These limits can be taken in different ways, and a possible choice is to 
take the $L^{-1}\rightarrow 0$ limit first, keeping temperature fixed, and then take $T\rightarrow 0$. 
We can also keep $L$ fixed and take $T\rightarrow 0$ first, and then $L^{-1}\rightarrow 0$. More generally we can 
take the limit $T,L^{-1}\rightarrow 0$ keeping  
\be
\label{eq:scaling}
TL^{\alpha}=\text{const.}
\ee
fixed, where $\alpha$ is a nonnegative real number. Generically, the constant has a noninteger mass dimension, for the practical implementation (e.g., for $\alpha=2$ below) one can fix $(T/m)(Lm)^\alpha$ to a dimensionless constant.
Notice that the consecutive limits mentioned above
correspond to the limiting cases $\alpha = 0$ and $\alpha = \infty$.
Here we will consider these two scaling trajectories plus 
$\alpha=1,2$ as we illustrate in Fig.~\ref{fig_scaling_diagram}. 
In this subsection we briefly discuss these different scaling trajectories
and partly describe the resulting physics using model calculations for the behavior at the emerging phase
transitions.

In the limiting trajectories $\alpha=0$ and $\alpha = \infty$ 
one has to perform the two limits consecutively. This is 
difficult to implement on the lattice but reveals interesting physics.
Therefore, we will use simple model calculations 
to illustrate characteristic features of the consecutive limits.
We will compare our numerical data 
to results for \textit{free one-dimensional fermions}: At low densities the behavior is governed by the small 
momentum-exchange between the particles, which for 
the O(3) model at hand are one-dimensional repulsive bosons.
At low momentum their behavior is universal
and given by free (spinless) fermions, see Sec.~\ref{sec_comparison}.

For the other two scaling trajectories, $\alpha=1$ and $\alpha=2$ the results will not differ much 
concerning the particle density, but are characterized by a different
behavior of the spin stiffness, which we will discuss in detail in Subsection~\ref{sec_qpt_stiff}.

\bigskip
\noindent
\underline{The scaling trajectory $\alpha=0$} \\
\\
The $\alpha=0$ trajectory corresponds to the consecutive limits
\[
L^{-1} \to 0 \text{ at fixed } T, \text{ then } T\to 0\;.
\]
The first limit of this sequence corresponds to the finite volume scaling at fixed nonzero temperature studied in 
Subsection~\ref{section_finite_volume}. This analysis has revealed a \textit{crossover} as seen from 
Fig.~\ref{fig_volumescaling}. Actually every fixed temperature possesses a specific 
crossover curve $n(\mu)_T$, which upon lowering the temperature $T$ to zero 
in the second limit might become steeper near some $\mu_c$ and turn into a genuine phase transition. 
To some extent this feature is seen in our data shown in Fig.~\ref{fig_tempscaling} below, where it is, however, overlaid
by the formation of ``condensation steps'' (see the discussion below).

\begin{figure}[b]
\includegraphics[width=\linewidth,type=pdf,ext=.pdf,read=.pdf]{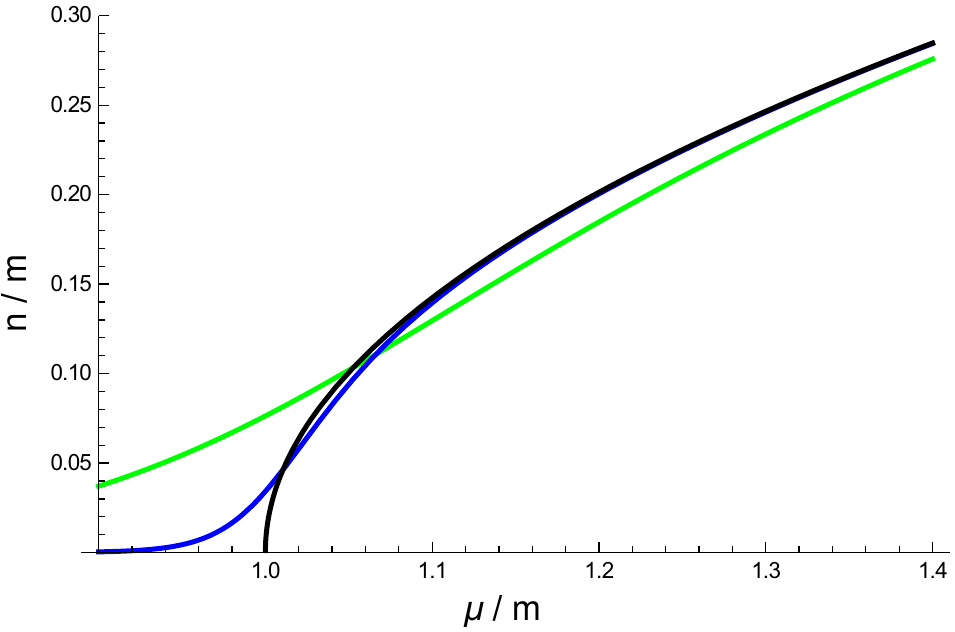}
\caption{Emergence of a phase transition in the limit $T \to 0$ for free one-dimensional fermions in an infinite volume. 
We show that the smooth curves for $T/m=1/10$ (green) and $T/m = 1/50$ (blue) approach 
the $T = 0$ square root behavior from Eq.~\eqref{eq_ferm_two}.
This is a model calculation for the scaling trajectory $\alpha=0$, especially its second limit $T\to 0$ at $L^{-1}=0$.}
\label{fig_ferm_root}
\end{figure}

The emergence of such a quantum phase transition as $T$ approaches 0 is of course a well known
feature which already appears in the simple model of free fermions in one spatial 
dimension: The density $n$ is given by the Fermi-Dirac 
integral $(L^{-1}=0)$,
\begin{align}
 n_f = \int_{-\infty}^\infty\!\frac{dk}{2\pi}\,\frac{1}{1+\exp\big((m+k^2/2m-\mu)/T\big)} \; ,
 \label{eq_ferm_one}
\end{align}
where we use a nonrelativistic dispersion which is sufficient for a first illustration. 
The density can be expressed by a polylogarithm Li$_{1/2}$, from which one obtains a square root in the 
$T \rightarrow 0$ limit,
\begin{align}
 \frac{n_f}{m} \to 
 &\:\frac{\sqrt{2}}{\pi}\sqrt{\mu/m-1}
 \cdot\Theta_{\text{Heaviside}}(\mu/m-1)\notag\\
 &\:\mbox{for} \quad  L^{-1}=0,\, T \to 0 \; .
 \label{eq_ferm_two}
\end{align}
The square root behavior, which is non-analytic at $\mu/m = 1$, is indeed reached only for $T = 0$ via a sequence
of analytic crossover-type curves as shown in Fig.~\ref{fig_ferm_root}.

\bigskip
\noindent
\underline{The scaling trajectory $\alpha=\infty$} \\
\\
We have already pointed out that our results for the density as a function of $\mu$ 
start to show ``condensation steps'' when the temperature $T$ is lowered at fixed $L$. Thus 
it is interesting to study also the scaling trajectory $\alpha = \infty$, which is defined by the consecutive limits

\[ 
T \to 0 \text{ at fixed }L, \text{ then }L^{-1} \to 0 \; .
\]

\noindent 
Our numerical results for the first one of these two limits are shown in Fig.~\ref{fig_tempscaling}. When 
lowering the temperature $T$ at fixed $L$ we observe the emergence of plateaus in the density where the
particle numbers are integers. The plateaus correspond to sectors with fixed particle number, and are smoothed out by the finite temperature (the width of the transition region being proportional to the temperature).  
Therefore, sharp \textit{steps} emerge in the $T\to 0$ limit at any fixed $L$.

\begin{figure}[b]
\includegraphics[width=\linewidth,type=pdf,ext=.pdf,read=.pdf]{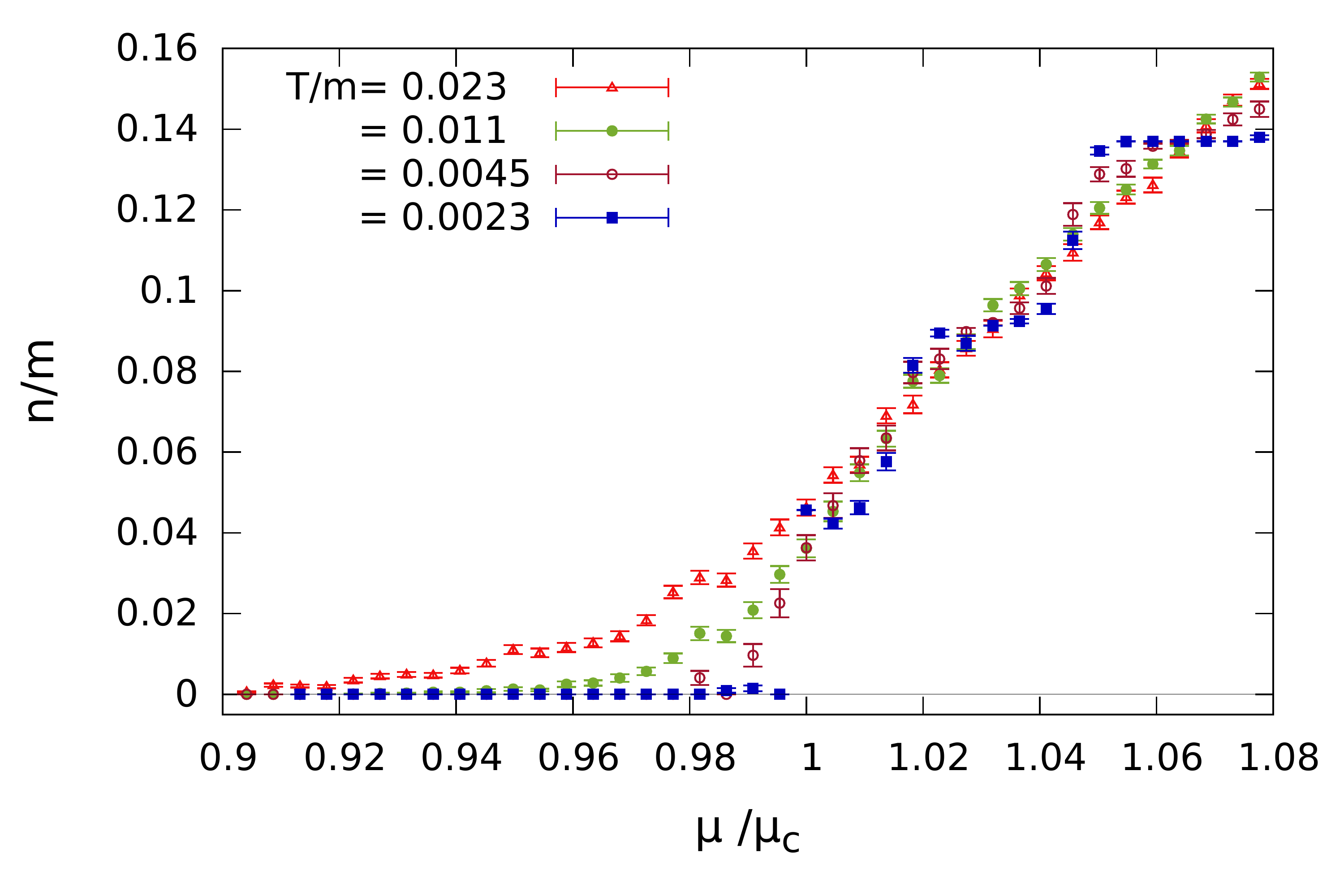}
\caption{Particle density $n$ in units of the mass 
as a function of $\mu$ for different temperatures at a fixed volume of $L = 22/m$
($J=1.3$, $N_s=100, N_t=$ $200,\,400,\,1000,\,2000$). When lowering the temperature one observes 
\cite{BruGaKloSu_2} the emergence of plateaus at integer values of the particle number, which for the normalized 
density $n/m$ used here corresponds to integer multiples of $0.046$. 
}
\label{fig_tempscaling}
\end{figure}

\begin{figure}
\hspace{0.3cm}\includegraphics[width=0.9\linewidth,type=pdf,ext=.pdf,read=.pdf]{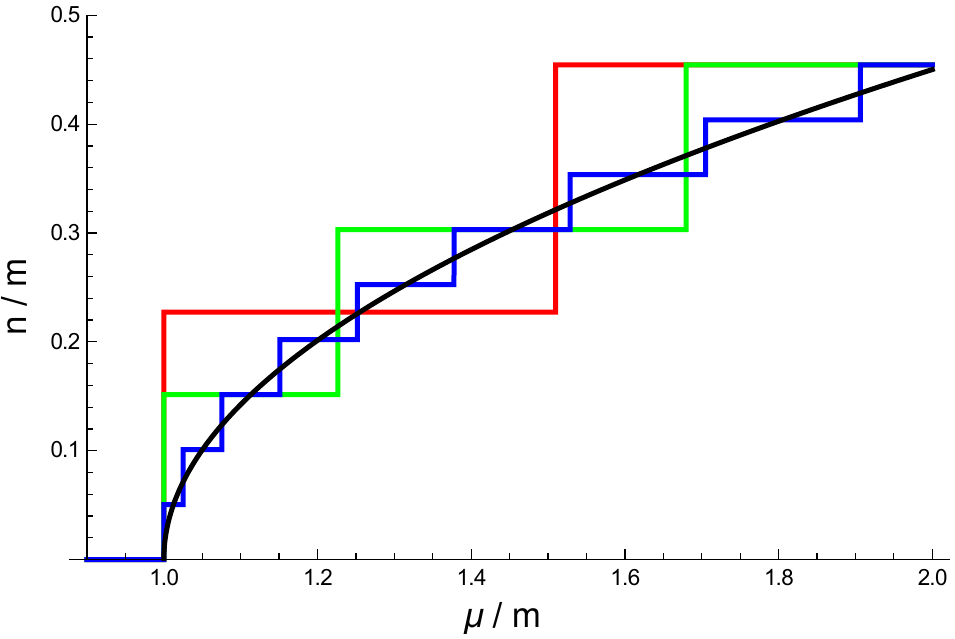}

\includegraphics[width=\linewidth,type=pdf,ext=.pdf,read=.pdf]{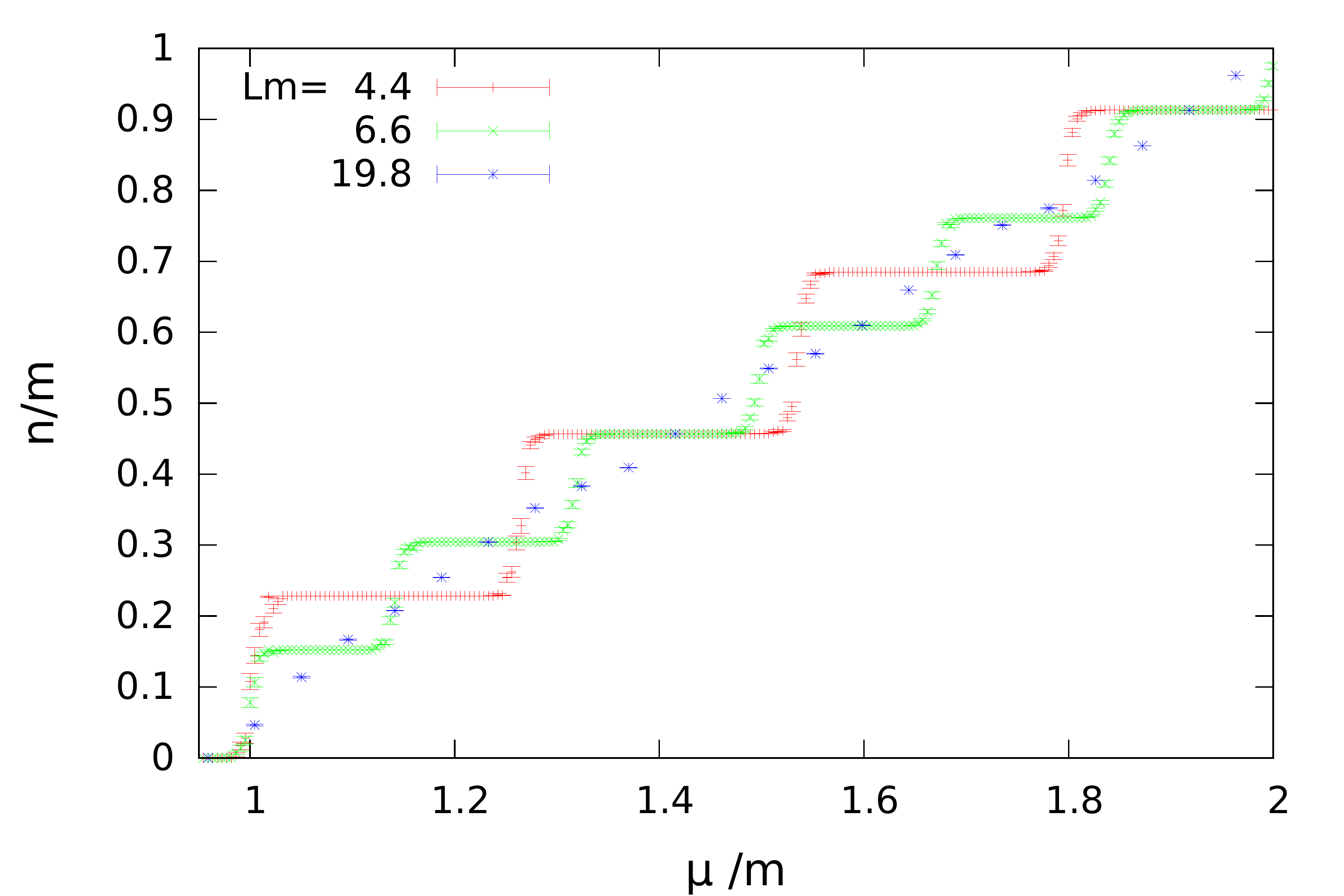}
\caption{Illustration of how the low temperature density with its steps at any finite volume 
approaches its infinite volume limit. Top: the Tonks-Girardeau gas similar to free fermions at zero temperature (see text) 
for $Lm=4.4,\,6.6,\,19.8$ (red, green, blue) being a model calculation for the scaling trajectory $\alpha=\infty$ (second limit). Bottom: Our results for the O(3) model at $T/m=0.0023$. For the different scales recall that free fermions approximate the O(3) model only at low densities.}
\label{fig_steps_larger_volumes}
\end{figure}

After considering $T =0$ at finite $L$ for following the $\alpha = \infty$ trajectory we need to 
perform the second limit $L^{-1} \to 0$. Also the behavior of this second limit can be illustrated 
in a one-dimensional fermion model, the Tonks-Girardeau limit of the Lieb-Liniger system, which we discuss in detail in 
Section~\ref{sec_comparison}. One finds that the minimal energy in each particle number sector 
$Q$ is $E_{\text{min}}^{(Q)} = \pi^2/(6L^2m)\cdot (Q^2-1)Q$.
Steps occur whenever 
$E_{\text{min}}^{(Q)}-Q\mu =E_{\text{min}}^{(Q-1)}-(Q-1)\mu$ and thus are located at 
$\mu_Q=m+\pi^2/2\cdot Q(Q-1)/(L^2m)$. The corresponding density 
$n/m=1/Lm\cdot\sum_{Q=1}^\infty \Theta_{\text{Heaviside}}(\mu-\mu_Q)$ is shown in the top panel of 
Fig.~\ref{fig_steps_larger_volumes}. It `oscillates' around its square root limit reaching it by ever smaller oscillations. The 
square root can also be seen analytically, since for large $Q$ the steps are at $\mu_Q/m=1+\pi^2/2\cdot(n/m)^2$.

The same behavior is seen in our numerical O(3) data shown in the bottom panel of Fig.~\ref{fig_steps_larger_volumes}, 
where, in the same fashion, `oscillations' around a limiting curve diminish with increasing $L$. 
However, this can only be done at low but nonzero temperatures and, as discussed above, this gives rise to a  
crossover instead of a phase transition.

\bigskip\clearpage
\noindent
\underline{The scaling trajectories $\alpha=1$ and $\alpha=2$:} 

\bigskip\noindent
In the scaling trajectories $\alpha=1$ and $\alpha=2$ we consider approaches to the limit $T \! =\!  0,\, L^{-1} \!=\! 0$ 
by sending $T$ and $L^{-1}$ to zero simultaneously. The trajectories differ in their functional relation 
between $T$ and $L$: 

\begin{align*} 
 \alpha=1: \quad  
 &T\to 0 \text{ and }L^{-1} \to 0\; \text{ with }T = L^{-1}\\
 \alpha=2: \quad  
 &T\to 0 \text{ and }L^{-1} \to 0\; \text{ with }T \propto (L^{-1})^2
\end{align*}

\begin{figure}[!b]
\includegraphics[width=\linewidth,type=pdf,ext=.pdf,read=.pdf]{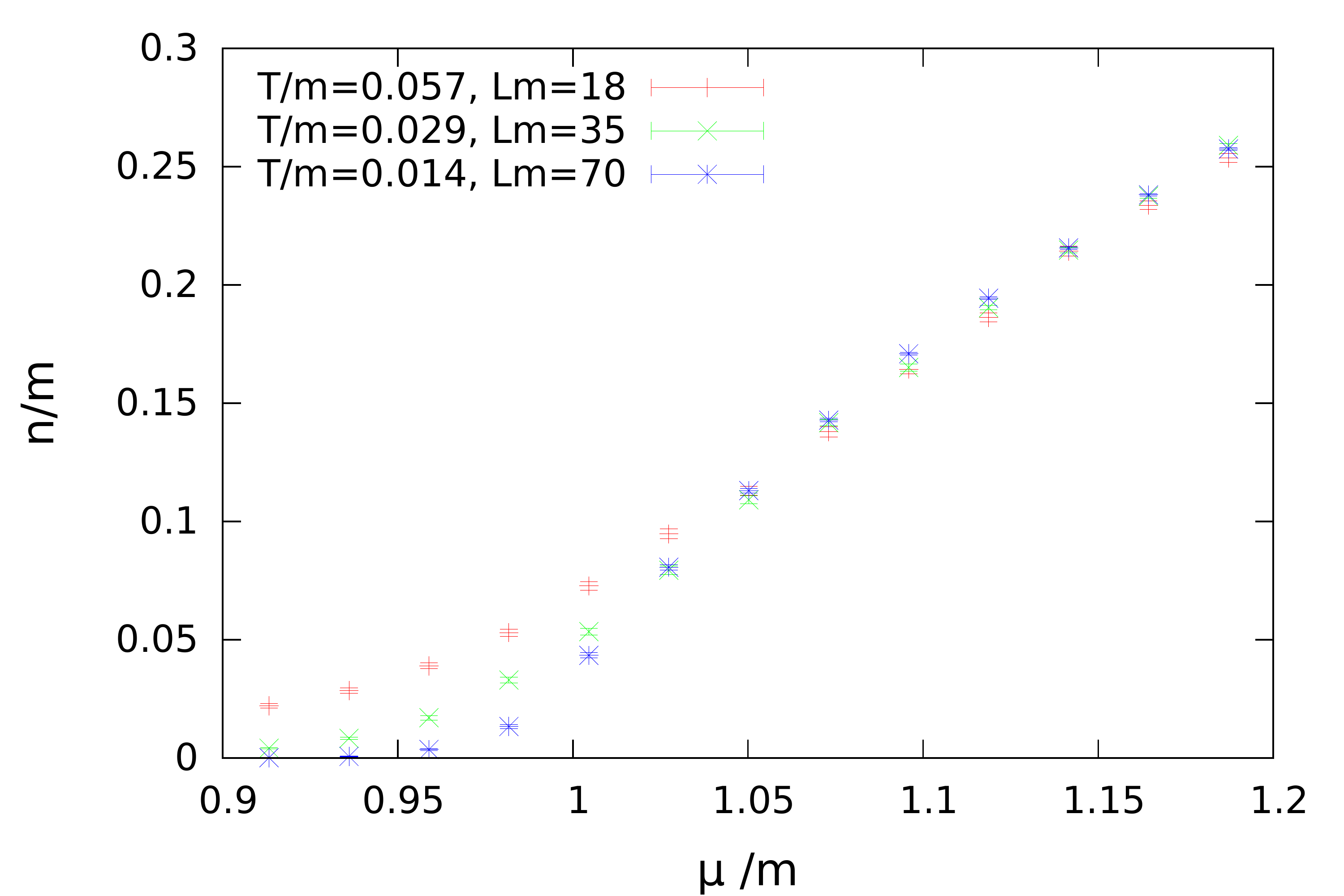}

\includegraphics[width=\linewidth,type=pdf,ext=.pdf,read=.pdf]{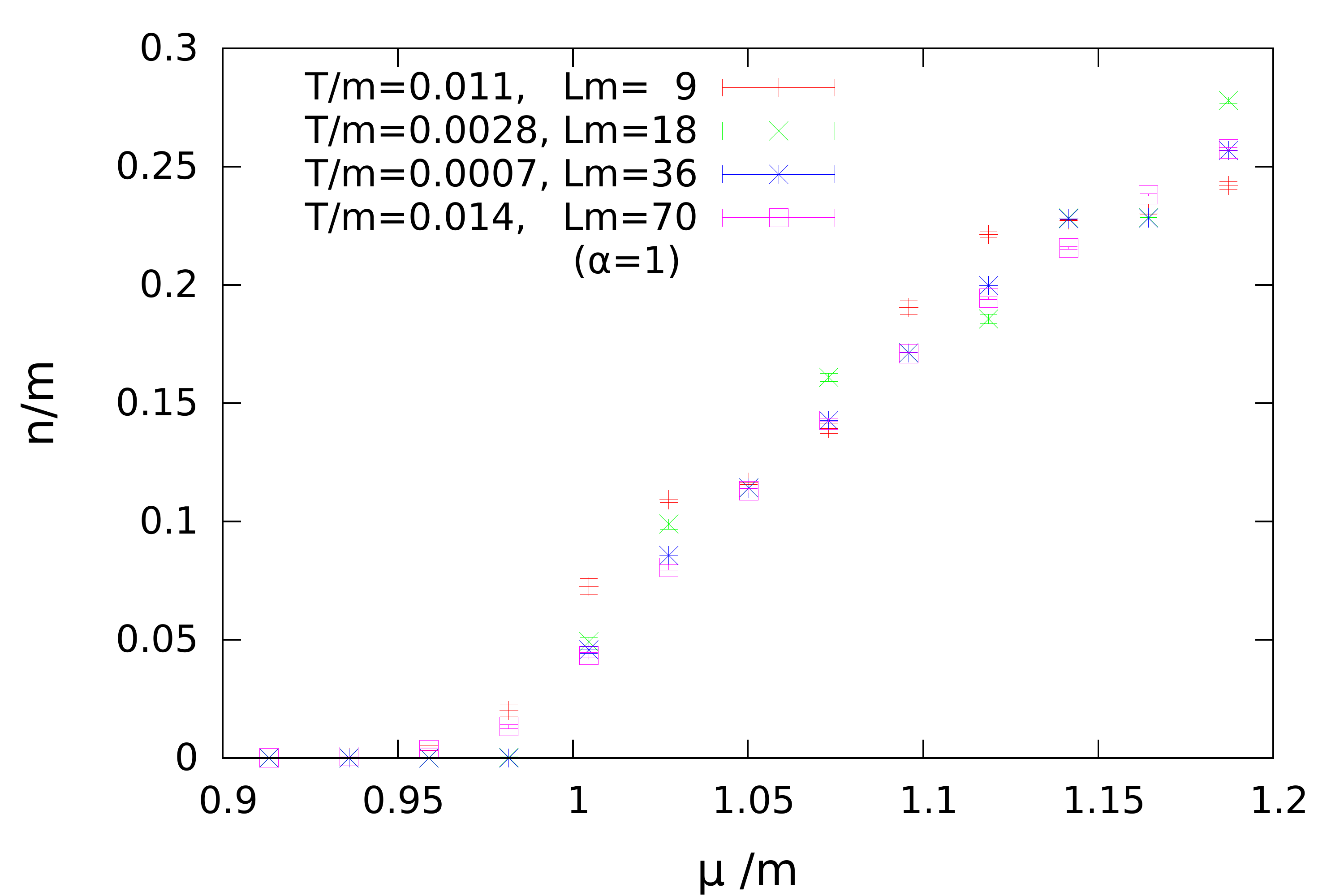}
\caption{Particle number density for the $\alpha=1$ (top) and $\alpha=2$ (bottom) scaling trajectories, both at $J=1.3$. 
For $\alpha=1$ we approach $T = 0$ using $L=1/T$ and show results for
$(N_t,N_s) \! =$ $(80,80), (160,160), (320,320)$. For $\alpha=2$ we scale with $Lm=0.94/\sqrt{T/m}$ using  
$(N_t,N_s)=(400,40), (1600,80), (6400,160)$ and show the lowest $T$ data from  $\alpha=1$ for comparison.}
 \label{fig_densities_both_scalings}
\end{figure}

\noindent
Obviously $\alpha=1$ corresponds to square lattices $N_t = N_s$, while $\alpha=2$ corresponds to time elongated 
lattices with $N_t \propto (N_s)^2$. To be specific, in our numerical simulations for  $\alpha=2$ we have used 
$N_t=(N_s/2)^2$ at $J=1.3$ ($am \sim 0.22$), which amounts to $T/m=0.88/(Lm)^2$ or $Lm=0.94/\sqrt{T/m}$. 

Fig.~\ref{fig_densities_both_scalings} shows our simulation results for these
two scaling trajectories. We show the particle density $n$ as a function of $\mu$ for $T\to 0$ and 
$L\to\infty$ simultaneously. One can see in both trajectories that $n$ vanishes for values of $\mu$ smaller than the 
threshold $m$, while it approaches certain nonzero values for $\mu>m$. This is an indication for a nonanalyticity 
developing at $\mu=m$ in the limit $T\to 0$, and thus for a \textit{quantum phase transition}. Whether it emerges in the 
form of a jump in $n$ (first order transition) or an infinite slope (second order) cannot be decided with our current data 
(and the common infinite volume limit $L\to\infty$ to distinguish the orders is part of the scaling to zero temperature). The 
agreement with analytic predictions for repulsive bosons presented in Sec.~\ref{sec_comparison} indicates a 
square root dependence on $\mu$ and thus a second order transition. 

Moreover, the data at lowest $T$/highest $L$ do not differ much between $\alpha=1$ and $\alpha=2$. This bulk quantity therefore does not seem too sensitive to the particular scaling trajectory in the limit $T\to 0$ and $L\to \infty$.

\subsection{Spin stiffness and dynamical critical exponent}
\label{sec_qpt_stiff}

\begin{figure}
\includegraphics[width=\linewidth,type=pdf,ext=.pdf,read=.pdf]{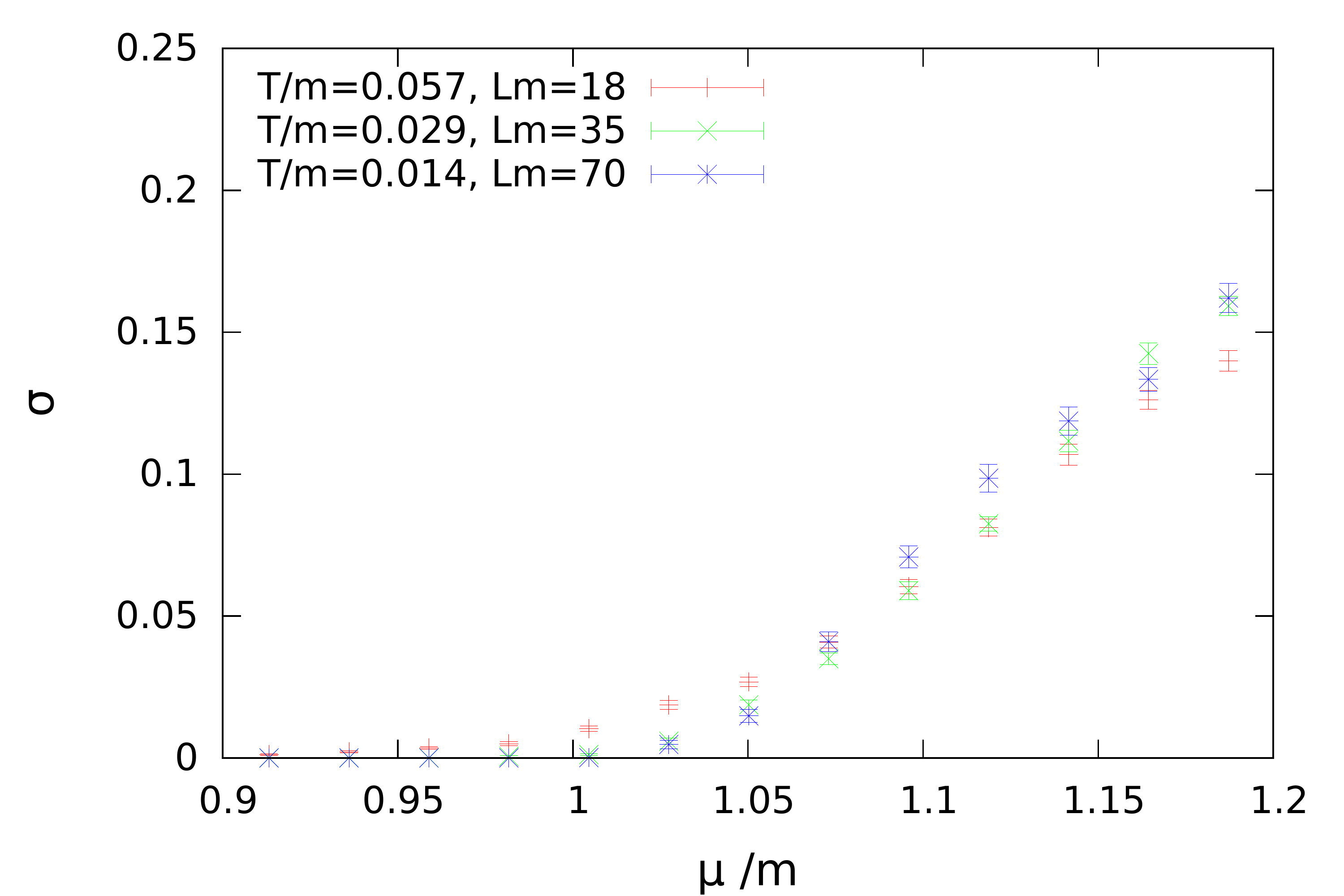}

\includegraphics[width=\linewidth,type=pdf,ext=.pdf,read=.pdf]{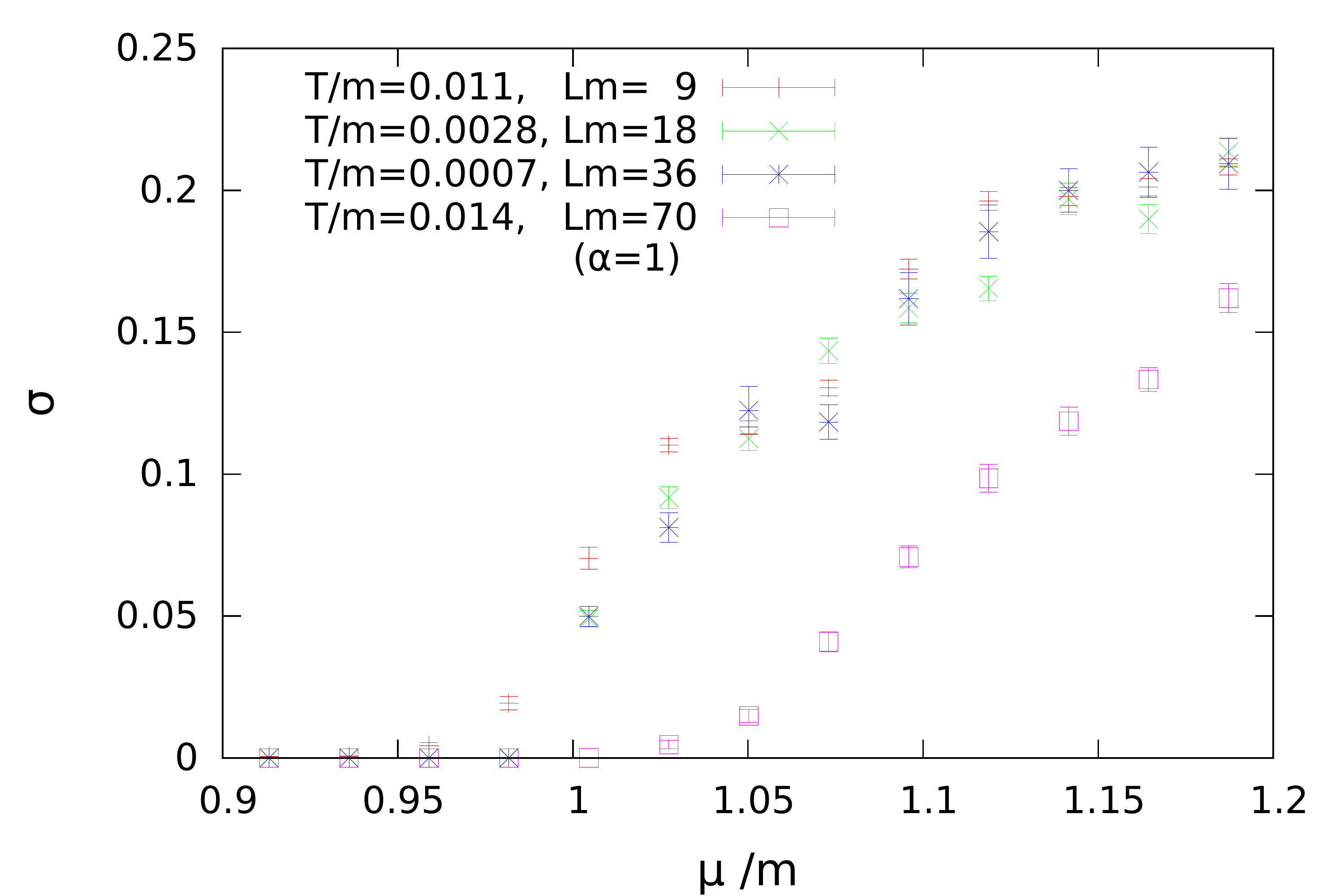}
\caption{Spin stiffness for the $\alpha = 1$ (top) and $\alpha = 2$ (bottom)  scaling trajectories 
(same parameters as in Fig.~\ref{fig_densities_both_scalings}).}
\label{fig_stiffness_both_scalings}
\end{figure}

\begin{figure}
 \includegraphics[width=\linewidth,type=pdf,ext=.pdf,read=.pdf]{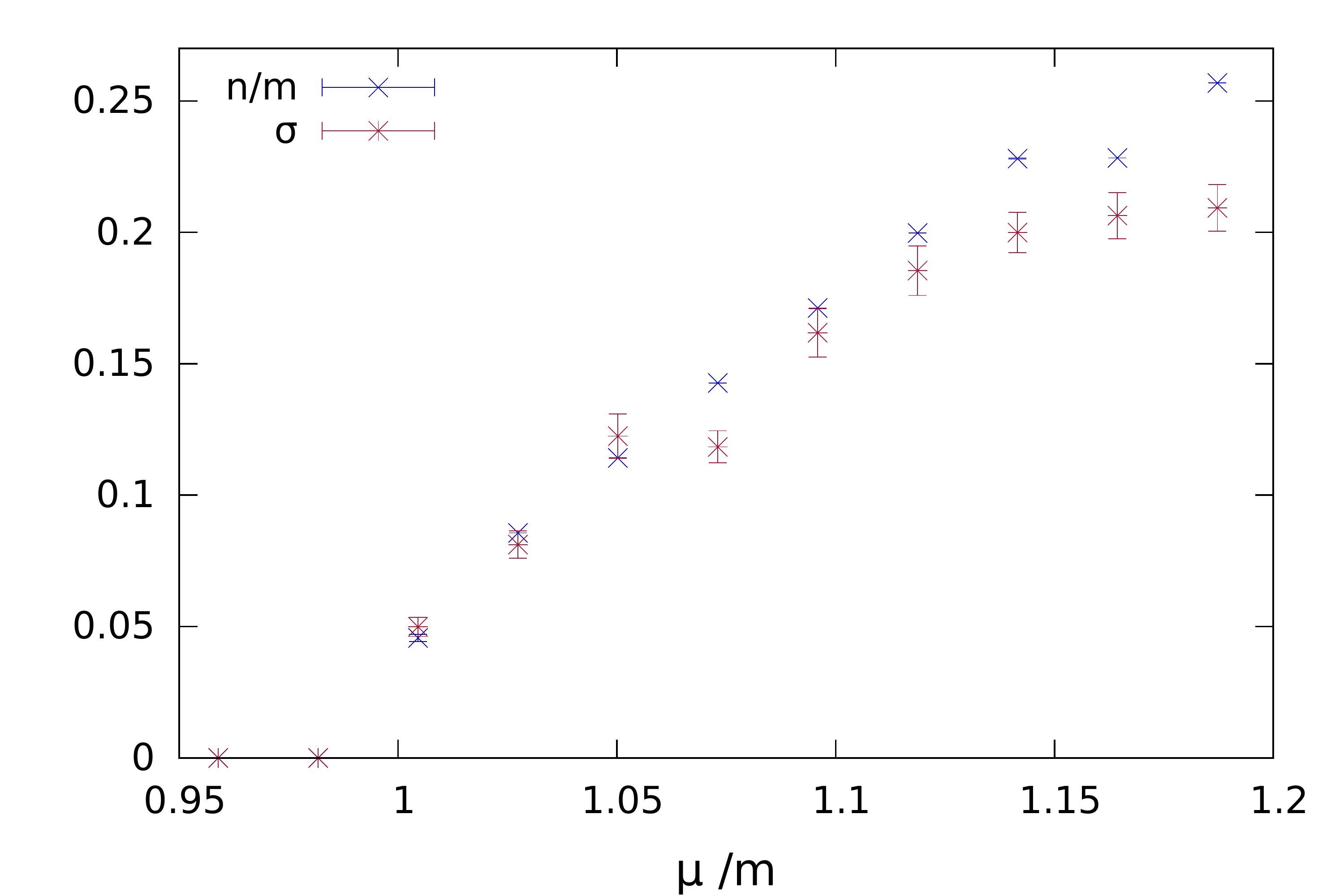}
 \caption{Agreement of stiffness and density (in mass units) close to zero temperature in a large volume $L\sim 1/\sqrt{T}$ ($\alpha = 2$, time elongated lattice): $T/m=0.0007$, $Lm=36$.}
 \label{fig_stiffness_vs_density}
\end{figure}

Fig.~\ref{fig_stiffness_both_scalings} shows our spin stiffness data for the two scaling trajectories $\alpha=1$ and $\alpha=2$. Qualitatively, the stiffness behaves like the density: it vanishes for $\mu<m$ and is nonzero for $\mu>m$. In contrast to the density, however, the stiffness shows a clear dependence on $\alpha$: for $\alpha=1$ it is \textit{significantly smaller} in the $\mu>m$ phase than for $\alpha=2$. In addition, the \textit{stiffness agrees with the density} (divided by mass) for $\alpha=2$ at small densities, as Fig.~\ref{fig_stiffness_vs_density} shows.
Our stiffness data do not display the signature of a BKT transition for $\mu>m$ in the spatial correlations suggested in \cite{falktin}. 

Let us now come to the interpretation of these data. It is known that the stiffness depends on the order of the limits 
$T\to 0$ and $L\to \infty$ \cite{Cazalilla-review}: when $L\to\infty$ is taken first (i.e. $\alpha=0$ scaling) the stiffness vanishes, whereas when 
$T\to 0$ is taken first (i.e. $\alpha=\infty$ scaling) it approaches the susceptibility of the ground state with respect to the same twist.

For $\alpha=0$ one keeps the temperature fixed (and thus the transition is a crossover), and then increases the spatial 
size to infinity. It is clear that in this way one looses all spatial correlations, the twist at the spatial boundary has no effect 
on the free energy for $L \rightarrow \infty$  
and the stiffness vanishes (the successive limit of zero temperature cannot change this any more). Our data for 
$\alpha=1$ suggest that this scaling trajectory is similar to $\alpha=0$ since the system size is growing too fast in 
comparison to $T\to 0$, such that spatial correlations are lost leading to a \textit{small stiffness}.  

For $\alpha=\infty$ one considers the zero temperature system at fixed volume and only afterwards the size is increased. At zero temperature one expects the ground state to determine the thermodynamic properties. In fact, the system at finite $L$ is gapped proportional to $1/L^2$. Therefore, the partition function in the $T\to 0$ limit is dominated by the ground state, and the free energy becomes the ground state energy (noting that the latter is $T$-independent, this is nothing but the third law of thermodynamics). As a consequence, the stiffness is the susceptibility of the ground state with respect to the spatial twist, as mentioned above. 

On top of that, we now invoke an argument from the Bethe ans\"atze of Sec.~\ref{sec_comparison}. One can easily convince oneself \footnote{The $I_i$ in Eq.~\eqref{eq_bethe_ansatz} come from closing the boundary and thus need to be shifted by $\varphi/(2\pi L)$. Moreover, it is sufficient to consider the nonrelativistic Bethe ansatz and thus replace $\theta\approx k/m \approx \sinh(\theta)$. Then the $I_i$-shift can be transfered to a uniform shift in the $k_i$, because the $\Delta$-term contains differences of $k$'s only and thus is unchanged.} that the (nonrelativistic) pseudo-momenta $k_i$ in the presence of a spatial twist $\varphi$ are shifted by $\varphi/L$. The ground state energy $\sum_{i=1}^Q k_i^2/(2m)$ receives no linear term in $\varphi$ since the total momentum $\sum k_i$ vanishes. The quadratic term is the constant $Q\cdot(\varphi/L)^2/(2m)$, and we immediately obtain
the stiffness as $\sti=Q/(Lm)=n/m$. This explains that the \textit{stiffness equals the density} for $\alpha=\infty$, for small 
densities. Our data for $\alpha=2$ suggest that this scaling trajectory is similar to $\alpha=\infty$ since 
the stiffness still equals the density, as seen in Fig.~\ref{fig_stiffness_vs_density}.

Finally, let us discuss the \textit{dynamical critical exponent}  \cite{Hohenberg:1977} $z$ that results from our findings. The correlation length near a second order phase transition is known to diverge. To achieve $T=0$ in the quantum phase transition one has to send the extent of the Euclidean time $\beta$ to infinity (just what we performed in the lattice simulations), and only at $\beta=\infty$ the phase transition occurs and the spatial correlation length $\xi$ is infinite. However, the system also has a typical correlation length $\xi_\tau$ in the Euclidean time direction. If nothing in the system distinguishes between space and Euclidean time, the two correlation lengths have to agree (which corresponds to $z=1$  
below). However, in general -- and in particular in the presence of $\mu$ which breaks the Euclidean symmetry by coupling to temporal 
components/winding numbers -- the two correlation lengths are related by 
$
 \xi_\tau \sim \xi^z
$ 
with $z$ the dynamical critical exponent \cite{Hohenberg:1977}. Note the similarity to our scaling trajectories $\beta\sim L^{\alpha}$ and indeed our stiffness data can be used to determine $z$ and the correlation length critical exponent $\nu$. 
For that we use finite size scaling of the free energy density  
\cite{Fisher:1989zza}, 
\begin{equation}
 f\sim\frac{1}{L\beta}\,g\big(\frac{\xi}{L},\frac{\xi^z}{\beta};\varphi\big)
\end{equation}
where the prefactor is the inverse volume (in 1$+$1 dimensions) making $f=F/L$ an intensive quantity and $g$ is a universal function of the spatial correlation length, the box sizes and the twist angle. Again, noninteger dimensions can be compensated by the corresponding powers of the mass.
Using the scaling $\xi\sim \delta^{-\nu}$ with the normalised distance to the critical value $\delta=\mu/m-1$, and the definition \eqref{eq_def_stiff} we get for the stiffness
\begin{equation}
 \sigma\sim L^{1-z}\,h(L^{1/\nu}\delta,TL^z)
\end{equation}
where $h$ is another universal function. 

\begin{figure}
 \includegraphics[width=\linewidth,type=pdf,ext=.pdf,read=.pdf]{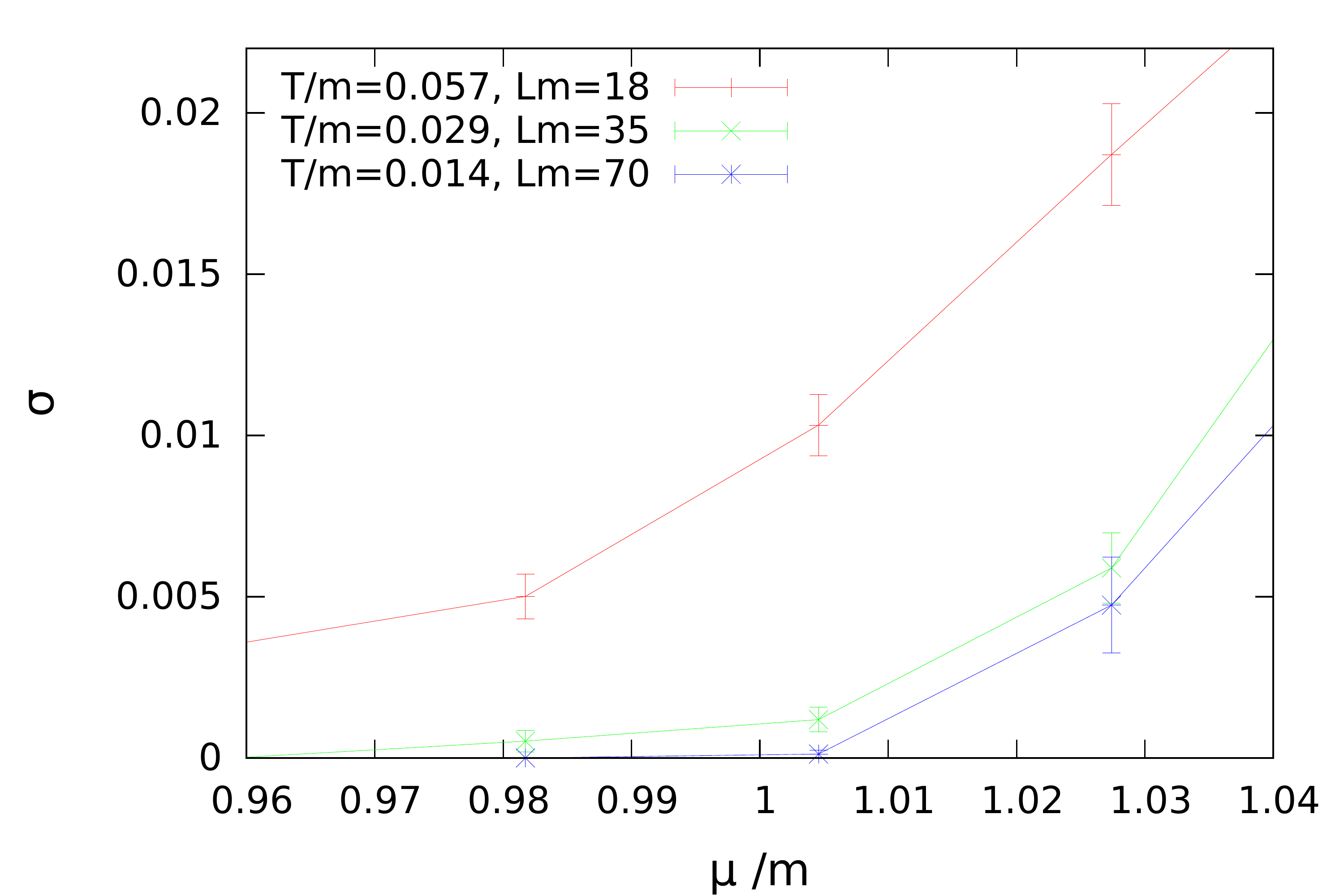}
 
 \includegraphics[width=\linewidth,type=pdf,ext=.pdf,read=.pdf]{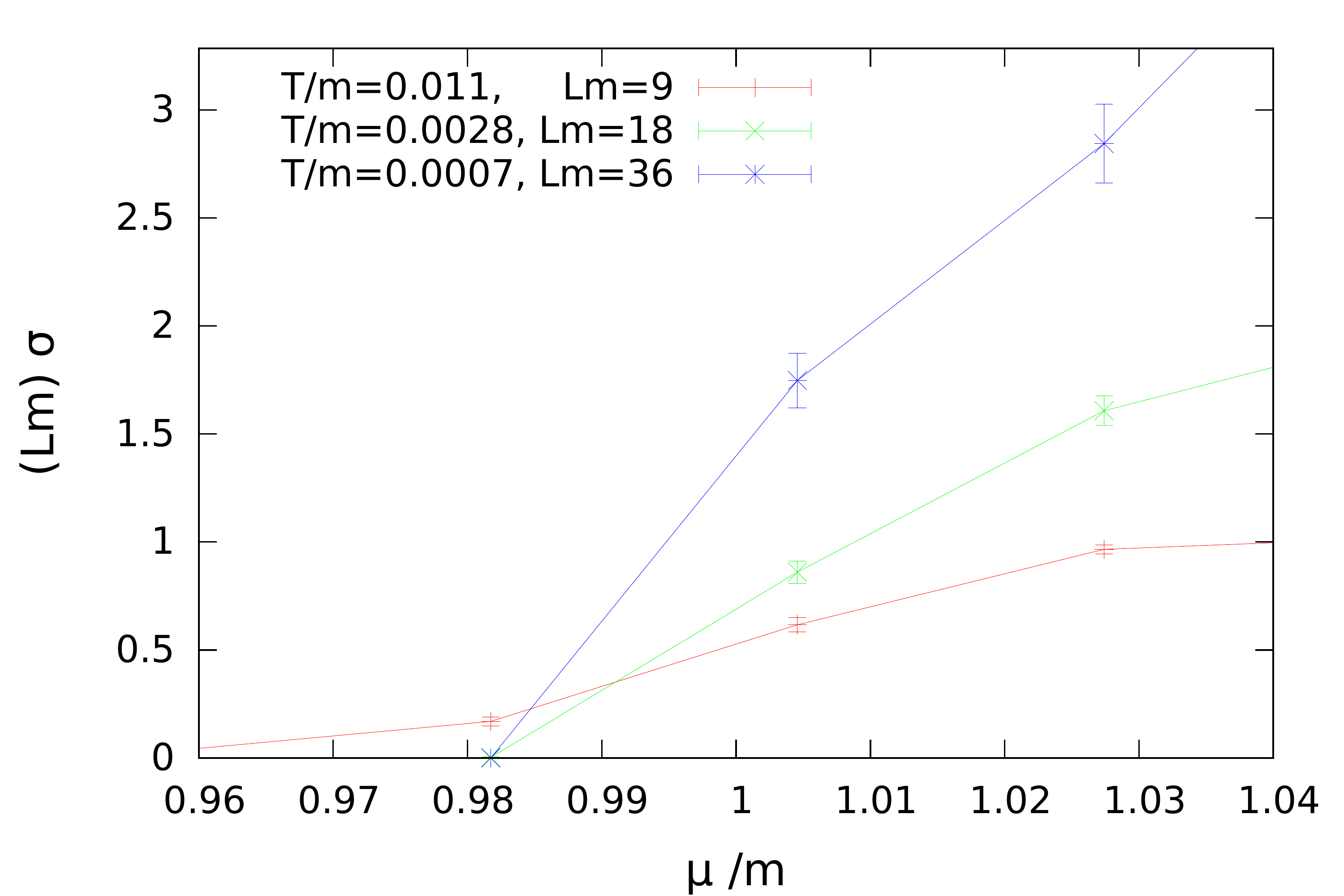}
  \caption{Numerical checks of statement (a) 
  for 
  exponents $z=1$ (top) and $z=2$ (bottom).
  For the lattice parameters see the caption of Fig.~\ref{fig_densities_both_scalings}.}
\label{fig:finite_size_first}
\end{figure}

For practical purposes one can derive two criteria \cite{vanOtterlo:1995}: 
\begin{align*}
 (a) &\text{ the curves }\mu\text{ vs.\ }L^{z-1}\sti 
 \text{ intersect at }\mu=m,\\
 &\text{ if }TL^z\text{ is kept constant}
\end{align*}
simply because at $\delta=0$ the length $L$ does not enter as argument of $h$ anymore
and
\begin{align*}
 (b) &\text{ the curves }L^{1/\nu}\delta\text{ vs.\ }L^{z-1}\sti
 \text{ collapse to a single curve,}\\ 
 &\text{ again if }TL^z\text{ is kept constant}
\end{align*}
Our data are collected at constant $TL^\alpha$ with $\alpha=1,2$ and thus we can ask whether $z=1$ or $z=2$ obey these statements, at least approximately. We will rescale all dimensionful quantites by the corresponding power of mass.

For statement (a) with $z=1$ we thus plot $\sti(\mu)$ keeping $TL$ constant, i.e., on the scaling trajectory $\alpha=1$, while for $z=2$ we plot $L^1\sti(\mu)$ keeping $TL^2$ constant, i.e., on the scaling trajectory $\alpha=2$, and see whether the data for different $L$'s intersect. As Fig.~\ref{fig:finite_size_first} shows, the curves from our numerical data intersect approximately for $z=2$ (albeit at $\mu$ slightly smaller than $m$), whereas for $z=1$ they certainly do not intersect. 

\begin{figure}
 \includegraphics[width=\linewidth,type=pdf,ext=.pdf,read=.pdf]{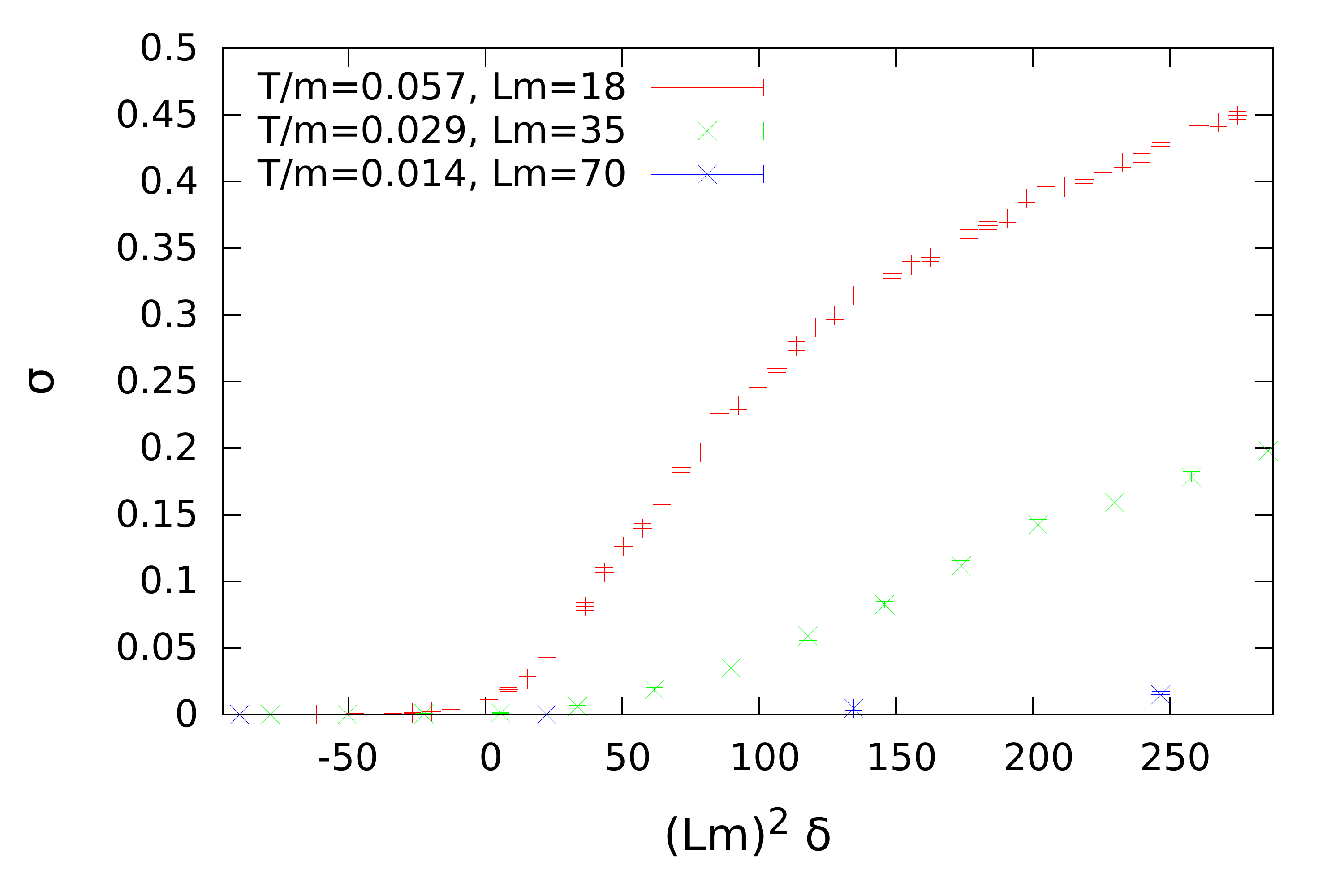}
 
 \includegraphics[width=\linewidth,type=pdf,ext=.pdf,read=.pdf]{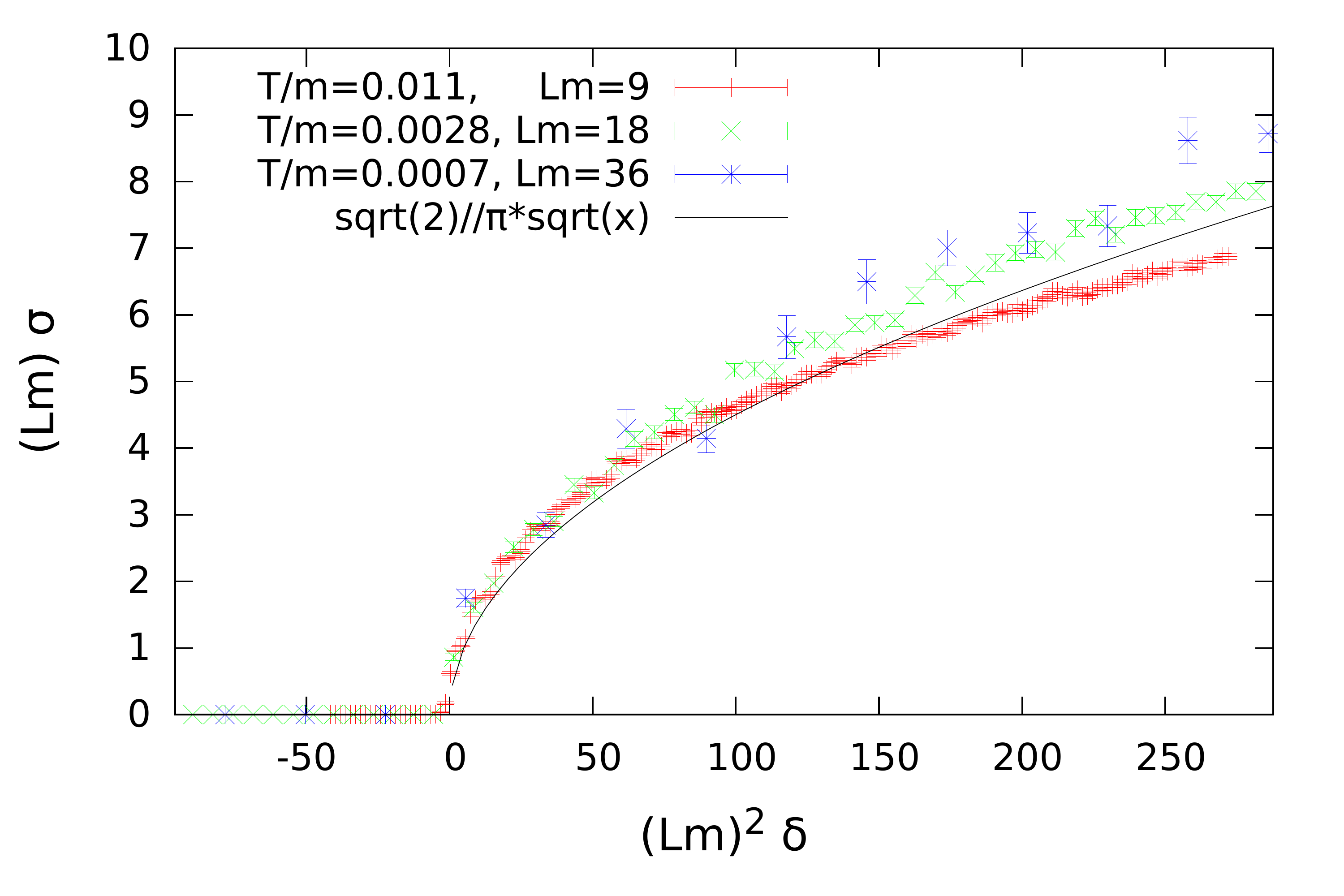}
 \caption{Numerical checks of statement (b) 
 for 
 $z=1$ (top) and $z=2$ (bottom) assuming $\nu=1/2$. In the bottom panel we also include the square root behavior in these variables.
 For the lattice parameters see the caption of Fig.~\ref{fig_densities_both_scalings}.}
\label{fig:finite_size_second}
\end{figure}

 \begin{figure}
 \includegraphics[width=\linewidth,type=pdf,ext=.pdf,read=.pdf]{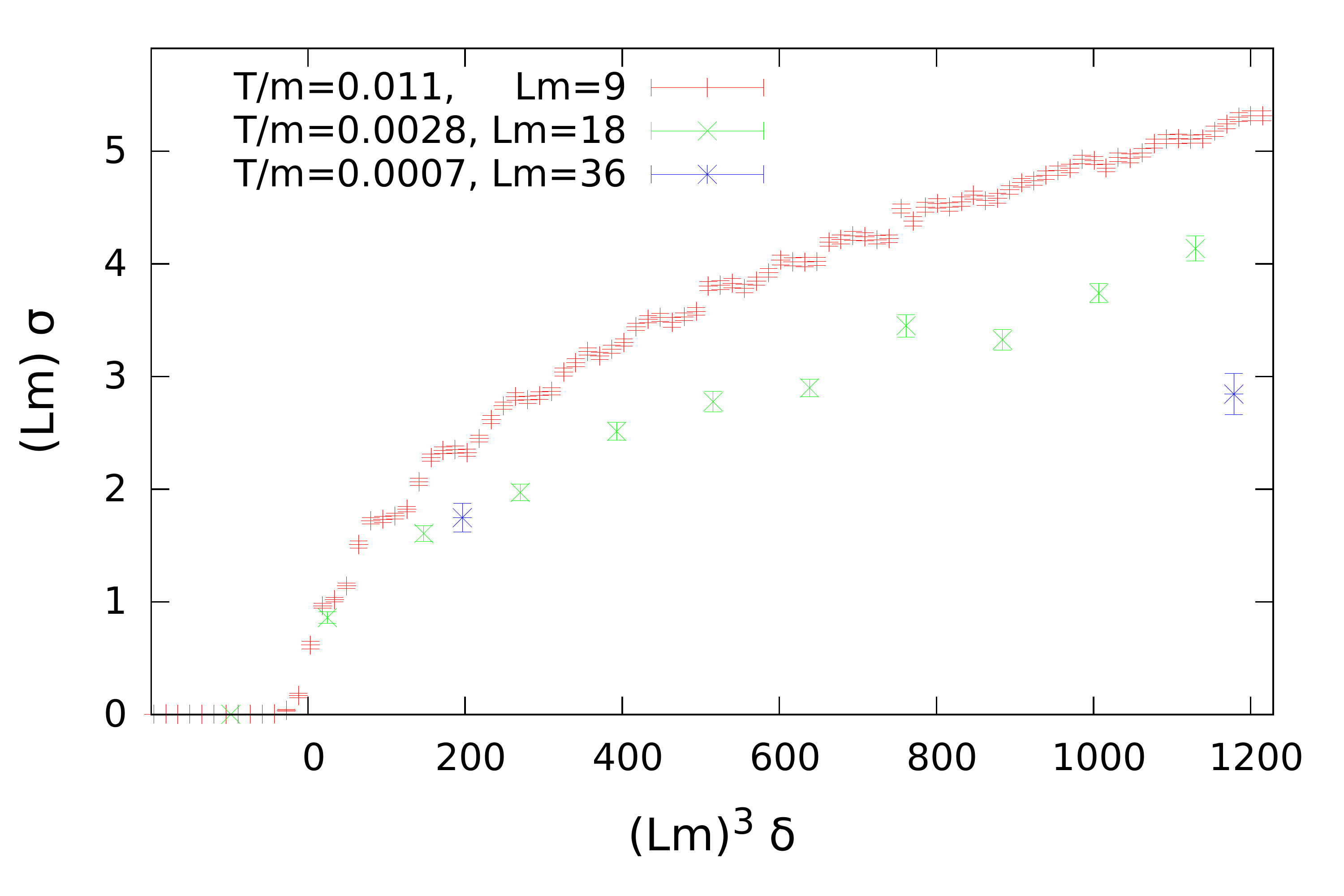}
 
 \includegraphics[width=\linewidth,type=pdf,ext=.pdf,read=.pdf]{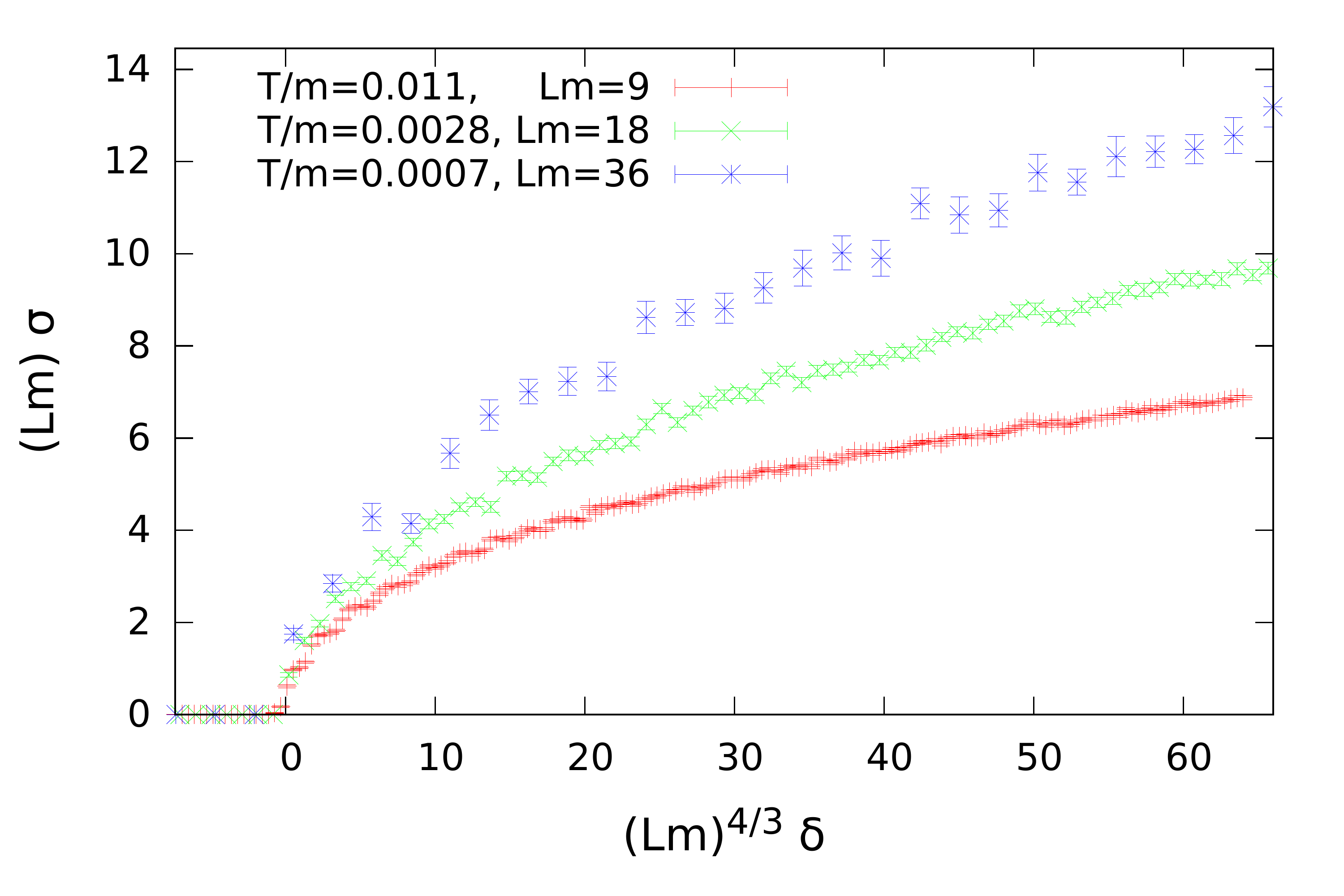}
 \caption{Numerical checks of statement (b) 
 assuming $z=2$ for $\nu=1/3$ (top) and $\nu=3/4$ (bottom).
 For the lattice parameters see the caption of Fig.~\ref{fig_densities_both_scalings}.}
\label{fig:finite_size_third}
\end{figure}

For statement (b) one has to deal with the appearance of another critical exponent $\nu$. In Fig.~\ref{fig:finite_size_second} we plot the obervables of this statement assuming the value $\nu=1/2$ for free fermions. Again a collapse to a single describes the situation much better for $z=2$. Finally, we can assume $z=2$ and check the value of $\nu$ away from $1/2$. Fig.~\ref{fig:finite_size_third} shows data proposing $\nu=1/3$ and $\nu=3/4$, for which the curves certainly do not collapse.
 
From this analysis we conclude that the critical exponents of the O(3) quantum phase transition at $\mu=m$ are consistent with $z=2$ and $\nu=1/2$, the values of free 
1d
fermions.
\enlargethispage{\baselineskip}

\section{Comparison to analytic results on repulsive bosons}
\label{sec_comparison}

In this section we will compare our density data to the density for 
systems of repulsive bosons in one spatial dimension. One of the best known examples is the one introduced by Lieb 
and Liniger (LL), 
where boson pairs interact via a Dirac delta function \cite{LiebLiniger}. The system can be solved in terms of plane 
waves picking up phase shifts $\delta$ when two bosons are interchanged. Such a Bethe ansatz \cite{Bethe:1931hc} 
works for nonrelativistic or relativistic bosons, without antibosons though. 

The phase shifts of the O(3) model are also known \cite{Zamolodchikov:1977nu} and those of ``isospin 2'',
\begin{align}
 \delta=-\arctan\frac{\pi}{2\theta}~,
\end{align}
with the relative rapidity,
\begin{align}
 \theta=\text{arsinh}(k/m) \; ,
\end{align}
govern the low density regime \cite{Hasenfratz:1990zz,BruGaKloSu_2}. 
We will thus use the Bethe ansatz equations with 
those phase shifts. Note that the O(3) wave functions are not plane waves anymore, but the Bethe ansatz is believed to 
give an exact result at low densities \cite{Sutherland}. 

At very low momenta the O(3) phase shifts agree with those of the LL model, actually all repulsive one-dimensional 
bosons are universal in the deep IR, where the precise UV shape of the interaction is not relevant and where $\delta(0)=-
\pi/2$ by Levinson's theorem \cite{Levinson}. Using only this value one arrives at the Tonks-Girardeau (TG) gas, which is 
the infinite coupling limit of the LL model \cite{Tonks,Girardeau}. Its eigenvalues are known to be that of free fermions
with the modification that the numbers $I_i$ in Eq.~\eqref{eq_bethe_ansatz} below are not always half-integers as is 
the case for antiperiodic fermions (which is not relevant for large $L$) and that the eigenfunctions of the TG gas are still 
symmetric under exchange of bosons. We have already shown that this results in the square root behavior 
\eqref{eq_ferm_two} of the particle density for $\mu\approx m$ at zero temperature and infinite volume 
(see also Eq.~\eqref{eq_square_root_again} below).

To be more precise, the relativistic Bethe ansatz reads
\begin{align}
 Lm \sinh(\theta_i) -2\sum_{\substack{j=1\\j\neq i}}^Q\Delta(\theta_j-\theta_i)=2\pi I_i~,
 \label{eq_bethe_ansatz}
\end{align}
where the $I_i$ are distinct, 
and half-integer/integer for even/odd total charge $Q$. In our conventions the dynamics enters via the phase shifts in 
\be\label{eq:Delta}
\Delta=\delta+\pi/2
\ee
with  $\lim_{k\rightarrow 0}\Delta=0$. This ansatz
works for ground states (with the choice $\{I_i\}=\{-(Q-1)/2,\ldots,(Q-1)/2\}$) and excited states at fixed 
particle number $Q$ in any finite volume $L$. The nonrelativistic Bethe ansatz can be obtained by approximating the 
rapiditites as $\theta\approx k/m$. Both are not too hard to solve numerically.

\begin{figure}
 \includegraphics[width=0.92\linewidth,type=pdf,ext=.pdf,read=.pdf]{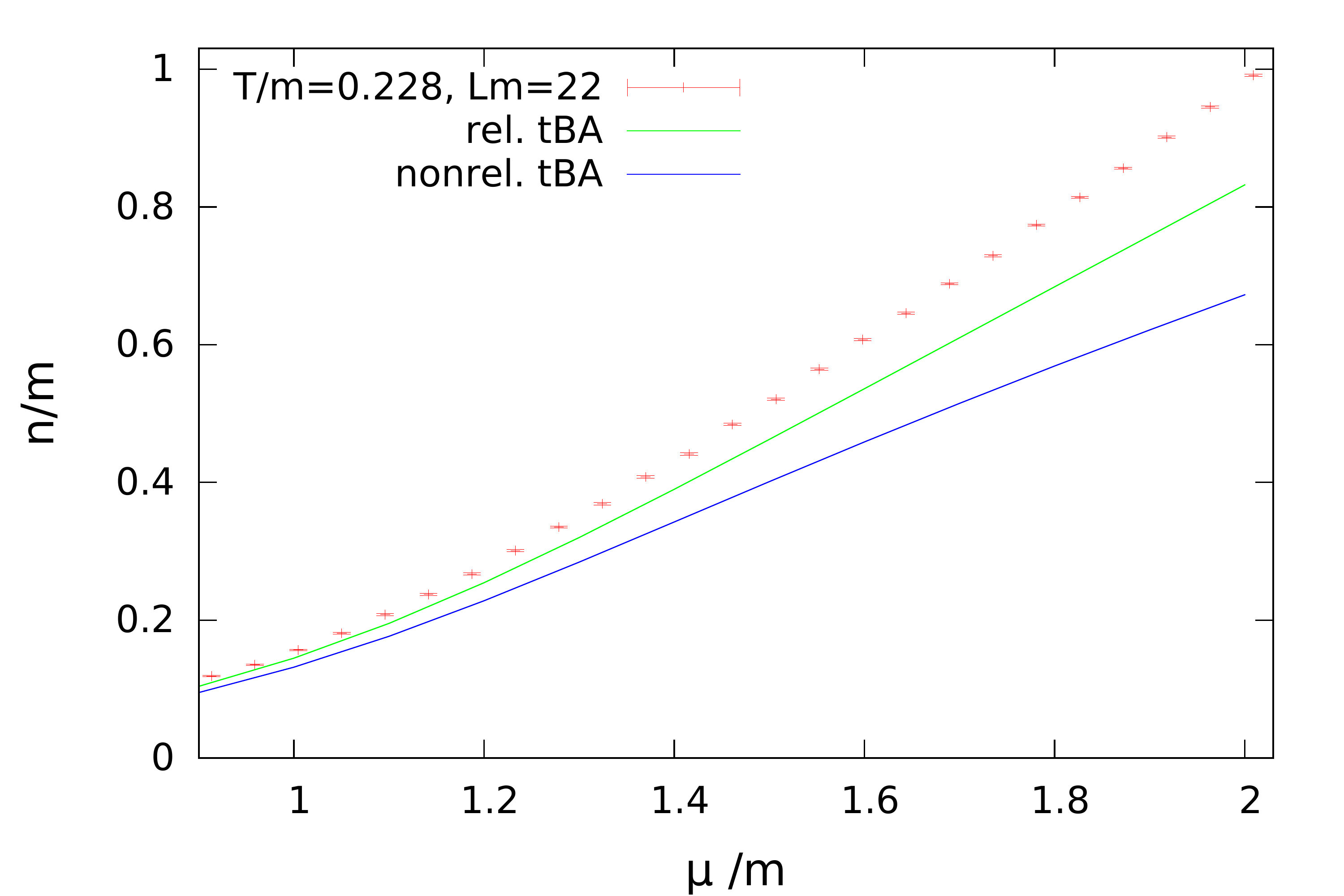}
 \includegraphics[width=0.92\linewidth,type=pdf,ext=.pdf,read=.pdf]{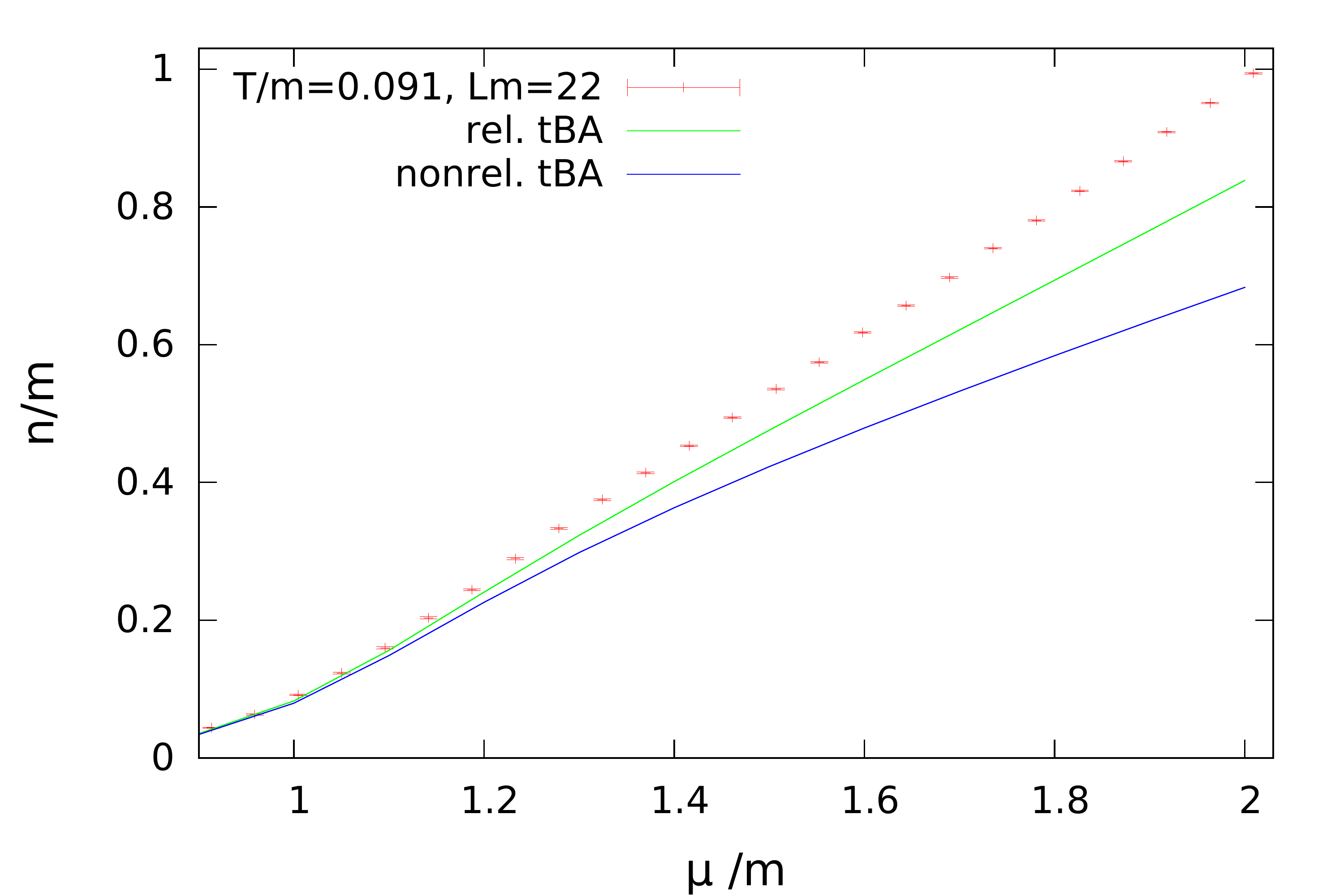}\\
 \includegraphics[width=0.92\linewidth,type=pdf,ext=.pdf,read=.pdf]{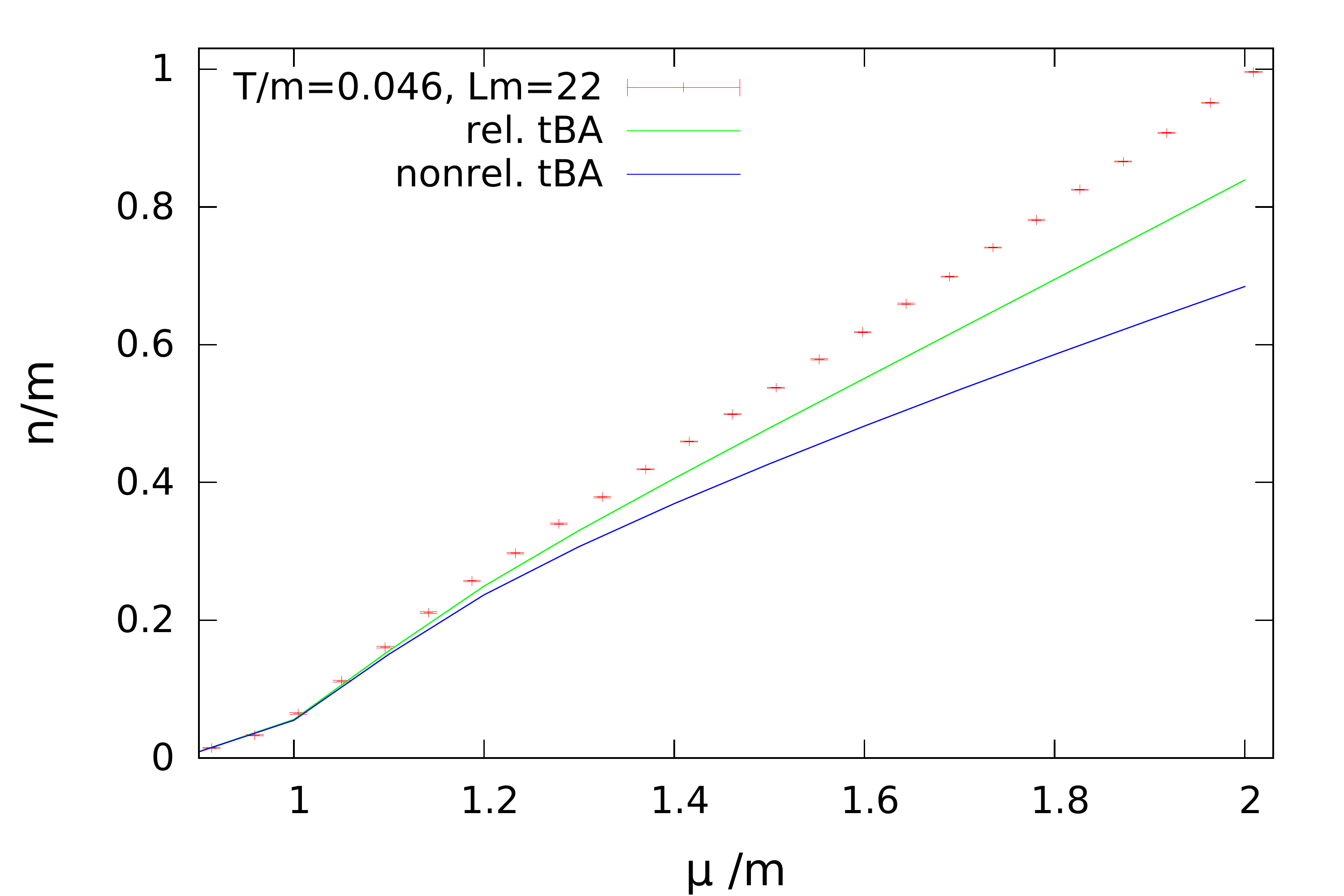}
 \includegraphics[width=0.92\linewidth,type=pdf,ext=.pdf,read=.pdf]{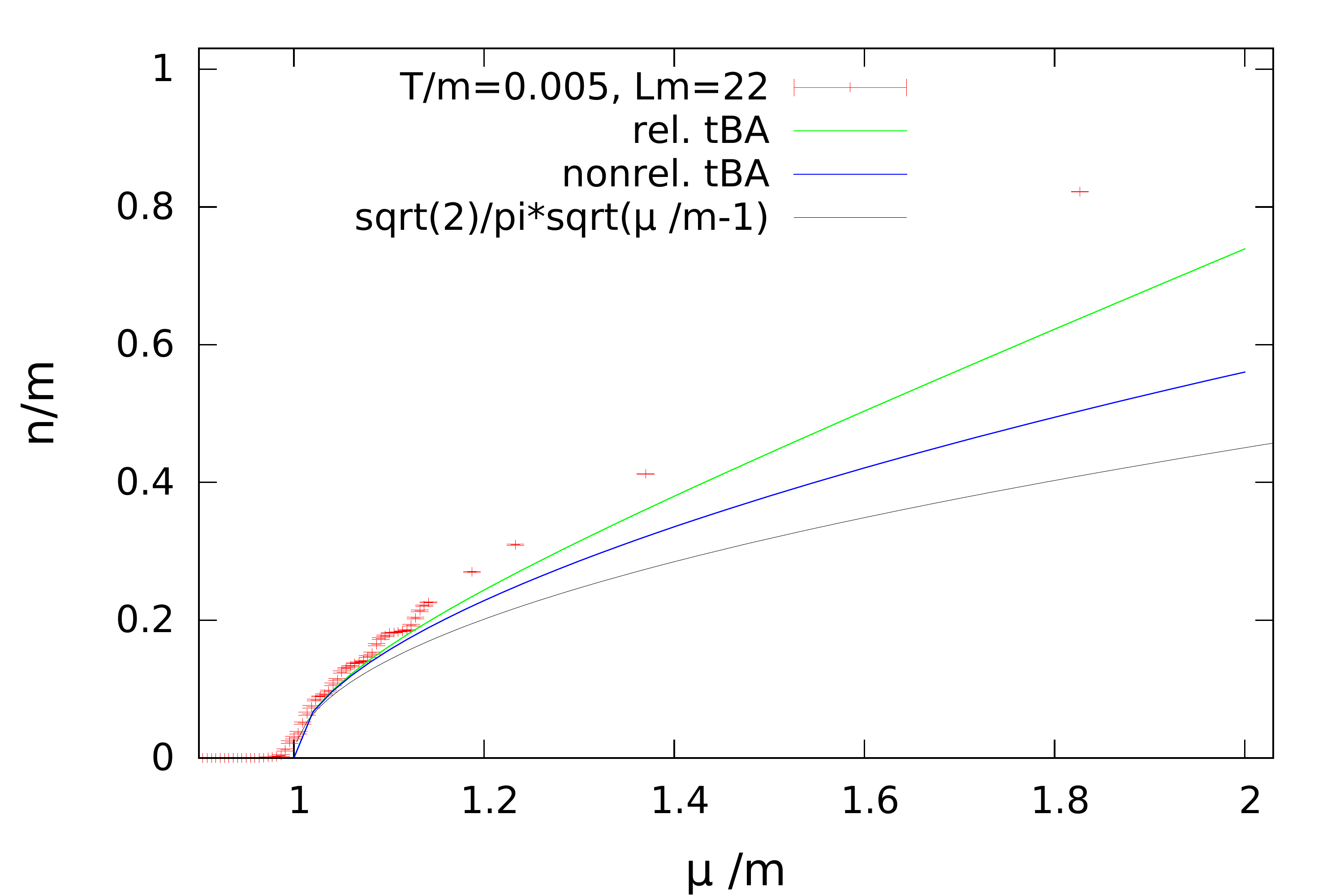}
 \caption{The density $n/m$ versus chemical potential $\mu/m$ at temperatures $T/m= 0.228,0.091,0.046,0.005$ from top to bottom. We compare the numerical evaluation 
 of the relativistic Bethe ansatz equation \eqref{eq:YYrel} (green curve) and the non-relativistic Bethe ansatz 
 equation of \eqref{eq:YYnonrel} (blue curve),  with the lattice data (red symbols, $J=1.3$,  $N_s=100$ and $N_t=20,\,50,\,100,\,1000$). 
 The black line in the bottom plot is the universal square-root behavior given in 
\eqref{eq_square_root_again}.}
\label{fig:latticecompare}
\end{figure}

Furthermore, Yang and Yang have derived a thermodynamic Bethe ansatz for the density $\rho(k)$ and energy density $
\epsilon(k)$ at nonzero temperatures and chemical potentials \cite{Yang:1968rm} in infinite volumes. The non-relativistic Yang-Yang equations give rise to
\begin{subequations}\label{eq:YYnonrel}
\begin{align}
 2\pi\rho(k)[1+e^{\epsilon(k)/T}]
 &=1+\int_{-\infty}^\infty\!d\tilde{k}\,
 \rho(\tilde{k})\,\Delta'(k-\tilde{k})~,\\
 \epsilon(k)
 &=\frac{k^2}{2m}+(m-\mu)-\frac{T}{\pi}\\
 &\times\!\!\int_{-\infty}^\infty\!\!\!d\tilde{k}\,
 \log\big(1+e^{-\epsilon(k)/T}\big)\,
 \Delta'(k-\tilde{k})\,.\notag
\end{align}
\end{subequations}
while, following similar arguments of \cite{Yang:1968rm}, and ignoring thermal pair production\footnote{Because of this, the equations are only exact at zero temperature. Nevertheless we will use them at finite, but small temperature to compare to lattice data.}, the relativistic Yang-Yang equations read 
\begin{subequations}\label{eq:YYrel}
\begin{align}
 2\pi\rho(\theta)\big(1+e^{\epsilon(\theta)/T}\big)
 &=m\cosh\theta+2\int_{-\Theta}^{\Theta}d\tilde\theta\;\rho(\theta)\Delta'(\theta-\tilde\theta)\;,\\
 \epsilon(\theta)
 &=m\cosh\theta-\mu
 -\frac{T}{\pi}\\
 &\times\!\int_{-\Theta}^{\Theta}\!\!d\tilde\theta\;\log(1+e^{-\epsilon/T})\Delta'(\tilde\theta) \; .\notag
 \end{align}
 \end{subequations}
 In both relativistic and non-relativistic equations, $\Delta'(x)=\frac{d\Delta(x)}{dx}$, where $\Delta$ is defined by \eqref{eq:Delta}.
Having solved this system one can extract the particle density as $n=\int_{-\infty}^\infty\!d\tilde{k}\,\rho(\tilde{k})$.

Finally, the zero temperature limit for the density is,
\begin{align}
 2\pi\rho(k)
 &=1+\int_{-K}^K\!d\tilde{k}\,
 \rho(\tilde{k})\,\Delta'(k-\tilde{k})~,\quad \rho(|k|>K)=0~,
\end{align}
where $K$ is the analogue of the Fermi momentum related to $\mu$ by $K=\sqrt{2}\,m\sqrt{\mu/m-1}$. 
Low densities amount to $\mu\approx m$ and thus $K\to 0$, for which the integration range shrinks, $\rho(0)=1/(2\pi)$ 
and $n=\rho(0)\,2K$ such that 
\begin{align}
 \frac{n}{m}
 =\frac{\sqrt{2}}{\pi}\, \sqrt{\mu/m-1}
 \qquad(T=0,\text{ small }n/m)~,
 \label{eq_square_root_again}
\end{align}
confirming once more the square root behavior for low densities.

We solve these integral equations numerically and compare to the lattice data in Fig. \ref{fig:latticecompare}. The agreement is fairly good. Note that these are parameter-free ans\"atze, so no fit is involved in the comparison. Let us repeat that the agreement is expected for low temperatures only, since Bethe ans\"atze do not contain antibosons and pair production.

\section{Summary}

In this paper we have presented thermodynamic lattice simulations of the asymptotically free two-dimensional 
O(3) model. The complex action problem of the conventional representation 
at finite chemical potential is overcome by using a representation in terms of dual variables, i.e., we simulated worldlines. In the O(3) model the mass of the particle triplet is generated dynamically, and in our simulations we have confirmed the expectation that at low temperatures a nonzero particle number (or charge) density occurs only when $\mu$ reaches the threshold given by the mass. Furthermore, in small volumes
several critical values of $\mu$ occur which correspond to integer particle numbers induced by $\mu$ 
\cite{BruGaKloSu_2} (for finite lattice couplings see the lattice phase diagram in Sec.~\ref{sec_lattice_diagram}).

A finite volume scaling analysis has revealed that 
at nonzero temperatures the particle density as a function of $\mu$ is regular, i.e., the transition is a crossover. 
The possibility of a phase transition at zero temperature has been analyzed through a 
simultaneous scaling of Euclidean time and space. Indeed, the data indicate a quantum phase transition at $\mu=m$, 
which, using the agreement with Bethe ans\"atze for one-dimensional repulsive bosons, should be of second order 
because the particle density follows the universal square root behavior. Further comparisons to these ans\"atze, 
including a nonzero 
temperature,  have shown that the O(3) model can indeed be described by these bosons. A more 
detailed analysis of this finding should include the continuum limit of our lattice data (which we believe will not change 
the results significantly, as we have observed for some observables at lattice coupling $J=1.4$), but is beyond the scope 
of the present study. 

We have also measured the spin stiffness which in the conventional representation 
is defined via twisted spatial boundary conditions and in the dual representation
measures spatial winding numbers of the worldlines. Based on the boson description at zero temperatures 
with only ground states, the stiffness is expected to be equal to the particle density. This is consistent with 
our low temperature data, when scaling $L$ such that $TL^2$ is constant, i.e., 
$\alpha=2$. However, when $L$ is larger in the zero 
temperature limit, e.g., when $TL$ is kept constant, i.e., $\alpha=1$, then this equality does not hold and the spin stiffness is significantly 
lower, which indicates a lost correlation between the spatial boundaries. From these findings we conclude that the dynamical 
critical exponent $z$ is close to $2$, in agreement with free fermions.

Although originally introduced for studying a potential BKT transition, our stiffness data have not indicated such a 
transition in the O(3) system at chemical potentials larger than the mass where it tends to be planar 
(as we have also confirmed).  

Besides higher precision, further studies in various directions would be useful: First of all, perturbation theory should 
match our lattice data at high $\mu$'s (keeping $a\mu$ small to avoid discretization effects). Secondly, measuring the vortex correlation functions or the topological susceptibility should shed light on the possibility of a BKT-like transition. Extensions to higher O(N) or CP(N-1) 
models as well as to 2+1 dimensions could be  done straightforwardly with the dual representation. 

The physically very interesting regimes of complex chemical potentials, where fractional instantons should occur (which in turn underly the resurgence program mentioned in the introduction), and at large theta angle, with the Haldane conjecture of a phase transition at theta angle equal to $\pi$, remain a challenge as 
the dual variables used here do not solve the sign problem in these situations.

\vskip5mm
\noindent
{\bf Acknowledgments:} 
We thank Quirin Hummel, Yannick Meurice, Lode Pollet, Klaus Richter, Juan Diego Urbina, Andreas Wipf and in particular Hans-Gerd Evertz for helpful discussions.
FB is supported by the DFG (BR 2872/6-1) and TK by the Austrian Science Fund, 
FWF, DK {\sl Hadrons in Vacuum, Nuclei, and Stars} (FWF DK W1203-N16).  
Furthermore this work is partly supported by the Austrian Science Fund FWF Grant.\ Nr.\ I 1452-N27 and by the DFG 
TR55, {\sl ''Hadron Properties from Lattice QCD''}. TS acknowledges the support of the DOE grant E-SC0013036. 

\appendix

\section{Critical $\mu$'s at strong coupling}
\label{app_strong_coupling_mus}

In this appendix we derive the critical values of the chemical potential in the strong coupling 
limit which we use for comparing to our numerical data in Figs. 2 and 3.
 
For small values of $J$, the configurations that dominate the partition function $Z$ of Eq.~\eqref{eq_dualZ} have minimal values of all dual variables,
\begin{align}
 \quad k_{x,\nu}=\overline{m}_{x,\nu}=0 \quad \forall x, \nu \quad \mbox{and}  \quad m_{x,1} = 0
 \quad \forall x~,
\end{align}
except the temporal component of the flux variable, which assumes a constant value
\begin{align}
 m_{x,2}=\dsc
 \quad \forall x~,
\end{align}
which amounts to $\dsc$ static particle world lines on each temporal bond. From Eq.~\eqref{eq_observables_dual_zero_densonly} it is clear that the resulting particle density is $n=\dsc$ (in lattice units). We restrict ourselves to positive $\mu$ and thus positive $\dsc$ for simplicity. 
These configurations obey the constraints in Eq.~\eqref{eq_dualZ} and result in the partition functions
\begin{align}
 Z_\dsc(J,\mu)=\left[\left(\frac{Je^{\mu}}{2}\right)^\dsc
 \,\frac{\Gamma(1/2)}{\Gamma(r+3/2)}\right]^{N_sN_t},
\end{align}
(where the factorials have cancelled against $\Gamma$-factors from the beta function) or grand potential densities
\begin{align}
 \Omega_\dsc(J,\mu)=-\dsc\ln\big(\frac{Je^{\mu}}{2}\big)+\ln\Gamma(\dsc+3/2) \; ,
\end{align}
up to an irrelevant additive constant.
Certain values of $\dsc$ yield the smallest grand potential $\Omega_\dsc$ depending on the values of $\mu$ and $J$. 
The neighboring values of $r$ take over, when the corresponding $\Omega_\dsc$'s become equal
\begin{align}
 \Omega_\dsc(J,\mu)=\Omega_{\dsc-1}(J,\mu) \quad \text{for some }\mu=\mu_\dsc~,
\end{align}
which gives the following critical chemical potentials $\mu_\dsc$ (in units of $a$),
\begin{align}
 e^{\mu_r}=\frac{2}{J}\,
 \frac{\Gamma(r+3/2)}{\Gamma(r+1/2)}\,,\qquad
 \mu_r=\ln((2\dsc+1)/J) \; .
\end{align}
Note further, that critical $\mu$'s inducing a density $n=\dsc$ induce a charge $Q=\dsc N_s$, thus 
\begin{align}
 \mu_{Q=\dsc N_s}=\ln((2\dsc+1)/J) \; ,
\end{align}
as used in Sec.~\ref{sec_lattice_diagram}.

\end{document}